\documentclass[12pt]{article}
\usepackage{graphicx}
\newcommand{\be}{\begin{equation}}
\newcommand{\ee}{\end{equation}}
\newcommand{\beq}{\begin{equation}}
\newcommand{\eeq}{\end{equation}}
\newcommand{\bea}{\begin{eqnarray}}
\newcommand{\eea}{\end{eqnarray}}

\begin{document}
\baselineskip=15.5pt
\pagestyle{plain}
\setcounter{page}{1}


\def\del{{\partial}}
\def\vev#1{\left\langle #1 \right\rangle}
\def\cn{{\cal N}}
\def\co{{\cal O}}
\newfont{\Bbb}{msbm10 scaled 1200}     
\newcommand{\mathbb}[1]{\mbox{\Bbb #1}}
\def\IC{{\mathbb C}}
\def\IR{{\mathbb R}}
\def\IZ{{\mathbb Z}}
\def\RP{{\bf RP}}
\def\CP{{\bf CP}}
\def\Poincare{{Poincar\'e }}
\def\tr{{\rm tr}}
\def\tp{{\tilde \Phi}}

\def\TL{\hfil$\displaystyle{##}$}
\def\TR{$\displaystyle{{}##}$\hfil}
\def\TC{\hfil$\displaystyle{##}$\hfil}
\def\TT{\hbox{##}}
\def\HLINE{\noalign{\vskip1\jot}\hline\noalign{\vskip1\jot}}
\def\seqalign#1#2{\vcenter{\openup1\jot
  \halign{\strut #1\cr #2 \cr}}}
\def\lbldef#1#2{\expandafter\gdef\csname #1\endcsname {#2}}
\def\eqn#1#2{\lbldef{#1}{(\ref{#1})}%
\begin{equation} #2 \label{#1} \end{equation}}
\def\eqalign#1{\vcenter{\openup1\jot
    \halign{\strut\span\TL & \span\TR\cr #1 \cr
   }}}
\def\eno#1{(\ref{#1})}
\def\href#1#2{#2}
\def\half{{1 \over 2}}

\def\ads{{\it AdS}}
\def\adsp{{\it AdS}$_{p+2}$}
\def\cft{{\it CFT}}

\newcommand{\ber}{\begin{eqnarray}}
\newcommand{\eer}{\end{eqnarray}}

\newcommand{\beqar}{\begin{eqnarray}}
\newcommand{\cN}{{\cal N}}
\newcommand{\cO}{{\cal O}}
\newcommand{\cA}{{\cal A}}
\newcommand{\cT}{{\cal T}}
\newcommand{\cF}{{\cal F}}
\newcommand{\cC}{{\cal C}}
\newcommand{\cR}{{\cal R}}
\newcommand{\cW}{{\cal W}}
\newcommand{\eeqar}{\end{eqnarray}}
\newcommand{\tht}{\thteta}
\newcommand{\lm}{\lambda}\newcommand{\Lm}{\Lambda}
\newcommand{\eps}{\epsilon}


\newcommand{\nonu}{\nonumber}
\newcommand{\oh}{\displaystyle{\frac{1}{2}}}
\newcommand{\dsl}
  {\kern.06em\hbox{\raise.15ex\hbox{$/$}\kern-.56em\hbox{$\partial$}}}
\newcommand{\id}{i\!\!\not\!\partial}
\newcommand{\as}{\not\!\! A}
\newcommand{\ps}{\not\! p}
\newcommand{\ks}{\not\! k}
\newcommand{\D}{{\cal{D}}}
\newcommand{\dv}{d^2x}
\newcommand{\Z}{{\cal Z}}
\newcommand{\N}{{\cal N}}
\newcommand{\Dsl}{\not\!\! D}
\newcommand{\Bsl}{\not\!\! B}
\newcommand{\Psl}{\not\!\! P}
\newcommand{\eeqarr}{\end{eqnarray}}
\newcommand{\ZZ}{{\rm \kern 0.275em Z \kern -0.92em Z}\;}

                                                                                                    
\def\del{{\delta^{\hbox{\sevenrm B}}}} \def\ex{{\hbox{\rm e}}}
\def\azb{A_{\bar z}} \def\az{A_z} \def\bzb{B_{\bar z}} \def\bz{B_z}
\def\czb{C_{\bar z}} \def\cz{C_z} \def\dzb{D_{\bar z}} \def\dz{D_z}
\def\im{{\hbox{\rm Im}}} \def\mod{{\hbox{\rm mod}}} \def\tr{{\hbox{\rm Tr}}}
\def\ch{{\hbox{\rm ch}}} \def\imp{{\hbox{\sevenrm Im}}}
\def\trp{{\hbox{\sevenrm Tr}}} \def\vol{{\hbox{\rm Vol}}}
\def\rl{\Lambda_{\hbox{\sevenrm R}}} \def\wl{\Lambda_{\hbox{\sevenrm W}}}
\def\fc{{\cal F}_{k+\cox}} \def\vev{vacuum expectation value}
\def\nodiv{\mid{\hbox{\hskip-7.8pt/}}}
\def\ie{{\em i.e.}}
\def\ie{\hbox{\it i.e.}}

\def\CC{{\mathchoice
{\rm C\mkern-8mu\vrule height1.45ex depth-.05ex
width.05em\mkern9mu\kern-.05em}
{\rm C\mkern-8mu\vrule height1.45ex depth-.05ex
width.05em\mkern9mu\kern-.05em}
{\rm C\mkern-8mu\vrule height1ex depth-.07ex
width.035em\mkern9mu\kern-.035em}
{\rm C\mkern-8mu\vrule height.65ex depth-.1ex
width.025em\mkern8mu\kern-.025em}}}
                                                                                                    
\def\RR{{\rm I\kern-1.6pt {\rm R}}}
\def\NN{{\rm I\!N}}
\def\ZZ{{\rm Z}\kern-3.8pt {\rm Z} \kern2pt}
\def\IB{\relax{\rm I\kern-.18em B}}
\def\ID{\relax{\rm I\kern-.18em D}}
\def\II{\relax{\rm I\kern-.18em I}}
\def\IP{\relax{\rm I\kern-.18em P}}
\newcommand{\CS}{{\scriptstyle {\rm CS}}}
\newcommand{\CSs}{{\scriptscriptstyle {\rm CS}}}
\newcommand{\rc}{\nonumber\\}
\newcommand{\bear}{\begin{eqnarray}}
\newcommand{\eear}{\end{eqnarray}}
\newcommand{\W}{{\cal W}}
\newcommand{\F}{{\cal F}}
\newcommand{\x}{{\cal O}}
\newcommand{\LL}{{\cal L}}
                                                                                                    
\def\mani{{\cal M}}
\def\calo{{\cal O}}
\def\calb{{\cal B}}
\def\calw{{\cal W}}
\def\calz{{\cal Z}}
\def\cald{{\cal D}}
\def\calc{{\cal C}}
\def\to{\rightarrow}
\def\ele{{\hbox{\sevenrm L}}}
\def\ere{{\hbox{\sevenrm R}}}
\def\zb{{\bar z}}
\def\wb{{\bar w}}
\def\nodiv{\mid{\hbox{\hskip-7.8pt/}}}
\def\menos{\hbox{\hskip-2.9pt}}
\def\dr{\dot R_}
\def\drr{\dot r_}
\def\ds{\dot s_}
\def\da{\dot A_}
\def\dga{\dot \gamma_}
\def\ga{\gamma_}
\def\dal{\dot\alpha_}
\def\al{\alpha_}
\def\cl{{closed}}
\def\cls{{closing}}
\def\vev{vacuum expectation value}
\def\tr{{\rm Tr}}
\def\to{\rightarrow}
\def\too{\longrightarrow}


\def\a{\alpha}
\def\b{\beta}
\def\c{\gamma}
\def\d{\delta}
\def\e{\epsilon}           
\def\f{\phi}               
\def\vf{\varphi}  \def\tvf{\tilde{\varphi}}
\def\vp{\varphi}
\def\g{\gamma}
\def\h{\eta}
\def\i{\iota}
\def\j{\psi}
\def\k{\kappa}                    
\def\l{\lambda}
\def\m{\mu}
\def\n{\nu}
\def\o{\omega}  \def\w{\omega}
\def\q{\theta}  \def\th{\theta}                  
\def\r{\rho}                                     
\def\s{\sigma}                                   
\def\t{\tau}
\def\u{\upsilon}
\def\x{\xi}
\def\z{\zeta}
\def\pt{\tilde{\varphi}}
\def\tt{\tilde{\theta}}
\def\lab{\label}  
\def\6{\partial}
\def\wg{\wedge}
\def\atanh{{\rm arctanh}}
\def\bpsi{\bar{\psi}}
\def\bt{\bar{\theta}}
\def\bvf{\bar{\varphi}}

%
                                                                                                    
\newfont{\namefont}{cmr10}
\newfont{\addfont}{cmti7 scaled 1440}
\newfont{\boldmathfont}{cmbx10}
\newfont{\headfontb}{cmbx10 scaled 1728}
\renewcommand{\theequation}{{\rm\thesection.\arabic{equation}}}

\font\cmss=cmss10 \font\cmsss=cmss10 at 7pt
\par\hfill KUL-TF-08/13, ITP-UU-08/39, SPIN-08/30

\begin{center}
{\LARGE{\bf Klebanov-Witten theory with massive dynamical flavors}}
\end{center}
\vskip 10pt
\begin{center}
{\large 
Francesco Bigazzi $^{a}$, Aldo L. Cotrone $^{b}$, Angel Paredes $^{c}$}\\
\end{center}
\vskip 10pt
\begin{center}
\textit{$^a$ Physique Th\'eorique et Math\'ematique and International Solvay
Institutes, Universit\'e Libre de Bruxelles; CP 231, B-1050
Bruxelles, Belgium.}\\
\textit{$^b$  Institute for theoretical physics, K.U. Leuven;
Celestijnenlaan 200D, B-3001 Leuven,
Belgium.}\\
\textit{$^c$ Institute for Theoretical Physics, Utrecht University; Leuvenlaan 4,
3584 CE Utrecht, The Netherlands.
}\\

{\small fbigazzi@ulb.ac.be, Aldo.Cotrone@fys.kuleuven.be, A.ParedesGalan@uu.nl}
\end{center}

\vspace{15pt}

\begin{center}
\textbf{Abstract}
\end{center}

\vspace{4pt}{\small \noindent 
We consider the addition of a large number of massive dynamical flavors to the Klebanov-Witten theory, the quiver gauge theory describing the low energy dynamics of $N_c$ D3-branes at the conifold singularity. Massive flavors are introduced by means of $N_f$ D7-branes which are holomorphically embedded and smeared along the transverse directions. After some general comments on the validity of the smearing procedure, we find the full backreacted supergravity solution corresponding to a particular class of massive embeddings. 
The solution depends on a running effective number of flavors, whose functional form follows from the smeared embedding. The running reflects the integrating in/out of massive degrees of freedom in the dual field theory as the energy scale is changed.
We study how the dynamics of the theory depends on the flavor parameters, mainly focusing on the static quark-antiquark potential. As expected, we find that the dynamical flavors tend to screen the static color charges.
}

\vfill

\newpage

\section{Introduction}
\setcounter{equation}{0}

The construction of string duals of gauge theories with dynamical flavors is a task of obvious interest. The gauge theories usually describe the low energy dynamics at the intersection of $N_c$ ``color'' and $N_f$ ``flavor'' D-branes and evaluating the backreaction of the full system is not easy in general. If the flavor branes are on top of each other, the supergravity equations of motion cannot be simply reduced to ordinary first order differential equations in a radial variable. Instead, they are partial differential equations in a number of variables given by the number of directions transverse to both the color and the flavor branes. 
For D3D7 models, this kind of scenarios have been studied in 
\cite{D3D7localized}. Other localized constructions were discussed in
\cite{otherlocalized}.

To avoid this complication one can consider 
simplified set-ups where the flavor branes are homogeneously smeared \cite{smearingtrick} in the transverse space. 
Whereas this restricts the possible ways of introducing fundamental 
matter (one requires that the symmetries of the unflavored theory
should be effectively unbroken), it gives a useful framework to analyze
physical features of the resulting flavored theories.
Moreover, the ``smearing trick'' is not only computationally helpful. 
As we are going to argue in the following section, 
it is in a sense a preferred choice if one wants to avoid singularities
and large string couplings near the position of localized sets of branes.

Following this prescription, 
the construction of string duals of flavored supersymmetric 
theories in the Veneziano limit $N_c,N_f\gg1$ with $N_f/N_c$ fixed, has been possible. In \cite{cnp1,cnp2,caceres} (resp. \cite{screening}) the string dual of a SQCD-like theory with massless (resp. massive) dynamical flavors was found. The setup is determined by D5-branes wrapped on compact (for the color branes) or non compact (for the flavor branes) two-cycles. The theory is the flavored version of the confining Chamseddine-Volkov-Maldacena-Nunez (CVMN) solution \cite{mn}. In  \cite{Benini:2006hh} a large number of massless flavors was added to the conifold conformal theory of Klebanov and Witten (KW) \cite{kw}.
In \cite{flavKS} the construction was extended to the confining Klebanov-Strassler \cite{ks} case. Other related setups and studies can be found in \cite{angeln2,beninis,ramallo3d,qgp}.

In this paper we focus on the flavored Klebanov-Witten model 
and extend the analysis of \cite{Benini:2006hh} by
considering the case where the dynamical flavors are massive and all with the same constituent mass $m_q$.
Fundamental flavor multiplets are added to the theory by means of D7-branes, 
which are wrapped on non-compact 4-cycles and holomorphically embedded in the background in order to preserve the ${\cal N}=1$ supersymmetry \cite{holomor}. 
The authors of  \cite{Benini:2006hh} found the supergravity solution 
generated by a ``massless'', smeared D3D7 system. This has a running dilaton and non trivial $F_1$ (sourced by the D7-branes) and $F_5$ RR fluxes. The near horizon geometry has no AdS factor and this translates in a broken conformal invariance in the dual field theory. The running dilaton blows up at a certain value of the radial coordinate corresponding to a field theory UV Landau pole. Moreover, the background has a curvature singularity when the dual gauge theory is in the far IR. 

In the following sections we will show how with massive
flavors one can find a IR regular solution. Of course, the latter will not alter the UV behavior of the field theory and the Landau pole will still be present. The flavor D7-branes are embedded in such a way that they reach a minimal radial distance $\rho_q$ from the bottom of the space. This distance is related to the mass parameter $m_q$, just as the field theory energy scale is related to the radius $\rho$. Energies larger (smaller) than $m_q$ map to radii larger (smaller) than $\rho_q$. 
The knowledge of the conifold geometry allows 
 to find the density distribution of the smeared flavor branes as a function of the radial coordinate. 

With the density distribution explicitly calculated, we are able to solve the supergravity equations of motion coupled to the $N_f$ D7-brane sources. The net result is interesting. As it was anticipated in \cite{Benini:2006hh}, the first order equations for the background fields following from the supersymmetric fermionic variations retain the same form as those in the massless case, modulo a substitution of $N_f$ with a running effective number of flavors, $N_f(\rho)$. The function $N_f(\rho)$ is related to the flavor density distribution mentioned above. This effective running of the number of flavors has a nice field theory interpretation. As we go towards energy scales much larger than the mass $m_q$ the theory resembles the massless one; at energies lower than $m_q$, the flavors can be integrated out and the theory looks like the unflavored one. We will see that the function $N_f(\rho)$, as well as the whole supergravity solution we will deduce from it, precisely reproduces this field theory expectation. The function $N_f(\rho)$ in fact has the shape of a smoothed out Heaviside step function $N_f\Theta(\rho-\rho_q)$. In contexts where extracting the function $N_f(\rho)$ from the smeared massive embedding could be technically difficult, it is useful to plug in the supergravity equations just the simple Heaviside function.  In this paper we will compare the results obtained by using the correct $N_f(\rho)$ with those deduced using the step approximation, finding that, in the flavored KW model at hand, the latter works quite well, at least at small $m_q$. This approximation was used to study novel effects of massive dynamical flavors in the confining CVMN background in \cite{screening}. 

Once the full flavored KW background has been found, one can use it to study how the strongly coupled dynamics of the dual gauge theory is affected by the dynamical flavors. One of the expected effects is the screening of the color charge. In this paper we use standard holographic techniques to extract the static potential between an external (i.e. extremely massive) quark-antiquark pair $\bar Q,Q$ probing our flavored gauge theory. The potential results to have a Coulomb-like shape and its behavior as a function of the sea quark parameters $N_f, m_q$ precisely accounts for the expected screening of the color charges. We also study the behavior of the screening length as well as that of the minimal quark-antiquark distance at which the $\bar Q Q$ bound state can decay into a pair of specific heavy-light bound states $\bar Q q+ \bar q Q$ (by popping out from the vacuum a dynamical quark-antiquark pair $\bar q, q$). Our analysis shows that the screening length is an increasing (resp. decreasing) function of $m_q$ (resp. $N_f$). The above mentioned minimal distance, which we call ``string breaking distance'' $L_{sb}$, is instead a decreasing function of both $N_f$ and $m_q$.

The paper is organized as follows. We start in Section \ref{sect: gensme} by reviewing the smearing technique, pointing out which are its limits of validity and its relevance. In Section \ref{sect: D7emb} we consider certain generalized massless and massive D7 embeddings on the conifold. Their smearing is considered in Section \ref{sect: sme} where the expression for their radial distribution density in a massive case is calculated. In Section \ref{sect: thesol} we find the supergravity background dual to the KW theory coupled to massive dynamical quarks and discuss the regimes where we can trust the solution. In Section \ref{sect: step} we rewrite the supergravity solution by using the Heaviside step function approximation. We then study, in Section \ref{sec: Wilson}, the dependence on the flavor parameters of the static quark-antiquark potential and of the screening and string breaking lengths, making a comparison between the results found using the ``true'' supergravity solution and those following from the step function approximation. In Appendix \ref{apporb} we review the orbifold origin of some D7 embeddings on the conifold. In Appendix \ref{app: integr} we study the dependence of the static quark-antiquark potential on a particular integration constant.


\section{Comments on the smearing technique} 
\label{sect: gensme}
\setcounter{equation}{0}

The technical trick we adopt to construct the string duals of gauge theories with a large number of dynamical flavors, consists in considering the flavor branes as homogeneously smeared along the transverse directions \cite{smearingtrick}. This trick allows one to find relatively simple solutions having the largest possible degree of symmetry and taking into account the full backreaction due to $N_c$ ``color'' and $N_f$ ``flavor'' D-branes.  Moreover, as we are going to argue, the trick is in a sense forced by the approximations used to find the full solution. In fact, the starting point on the low energy string side is an action of the form \cite{km}
\beq
S= S_{II}+ \sum_{i=1}^{N_f} [S_{i\,BI}+ S_{i\,WZ}]\,,
\label{BIWZ}
\eeq
where the first term is the (IIB or IIA) bulk supergravity action and the remaining ones are the Dirac-Born-Infeld and Wess-Zumino part of the flavor brane action, with the flavor branes embedded in the background. One usually first solves for the embedding equations and then for the remaining bulk equations on-shell. 

Let us now consider $N_f$ {\it coincident} flavor
branes, such that the corresponding flavor symmetry group is $U(N_f)$. In this case one expects corrections to the DBI action coming from the fact that an open string can end on the $N_f$ branes, so that the effective coupling is $g_s N_f$. In the quenched approximation $N_f\ll N_c$ and the coupling is small, but if $N_f=O(N_c)$ the string theory would be strongly coupled.
The DBI action was shown to appropriately describe the $g_sN_f$-leading order dynamics of open strings in a generic background \cite{Callan:1986bc}. Thus, if $g_sN_f$ is order one, the corrections to the DBI can be large.
This fact has a precise analogue in field theory in the Veneziano limit \cite{Veneziano:1976wm}. At any given order in $N_c$, the insertion of $n$ quark loops (``windows'') is governed by the parameter $(g^2_{YM}N_f)^n$ and can be done perturbatively only for small $g^2_{YM}N_f$. If the latter is not small, the sum over any number of windows must be performed non-perturbatively.
In the brane language, the ``one window graph'' corresponds to the DBI contribution, which is then a good approximation only for small  $g_sN_f$.

Crucially, {\emph{in the smeared setup}} the effective coupling $g_sN_f$ is further suppressed.\footnote{We are grateful to Carlos N\'u\~nez for this crucial observation and for his relevant contributions to the discussion presented in this Section.} Due to the smearing, the flavor symmetry $U(N_f)$ is generically broken to $U(1)^{N_f}$ and the distance between two generic flavor branes is large (in string units) in the supergravity approximation. 
Now, a typical string process will involve a space-time region of size of order one in $\sqrt{\alpha'}$ units: in such a region, thus, only a small fraction of the $N_f$ smeared flavor branes will be available for the process. If we denote as $R$ the typical radius of an internal dimension of the geometry in $\sqrt{\alpha'}$ units, the number of flavor branes involved in a typical process will be of order $N_f/R^d$, where $d$ is the codimension of the flavor branes (the number of dimensions involved in the smearing).
Since in the supergravity approximation $R\gg 1$, the effective coupling to the flavor branes $(g_s N_f)/R^d$ will typically
 be small even if $N_f = O(N_c)$.
Thus, in cases as \cite{cnp1}-\cite{screening},\cite{angeln2},\cite{ramallo3d}
 where the supergravity regime can be attained with no restrictions on $N_f/N_c$, the use of the DBI action is still justified also for $g_s N_f$ of order one.
For the case considered in this paper, nevertheless, we will show in  section \ref{ricci}
that the validity of the supergravity approximation implies $N_f \ll N_c$ and that the validity of the DBI
does not impose any further restriction.

A more detailed discussion of 
these issues will be presented in \cite{carlosetal}.
\section{D7 embeddings on the conifold}
\label{sect: D7emb}
\setcounter{equation}{0}

The low energy dynamics of $N_c$ D3-branes at the conifold singularity 
\be
z_1z_2=z_3z_4\,,
\ee
where the $z_i$ are complex coordinates, is described by an ${\cal N}=1$ superconformal quiver gauge theory with gauge group $SU(N_c)\times SU(N_c)$ and bifundamental matter fields $A_1, A_2$ and $B_1, B_2$ transforming respectively in the ({\bf $N_c,\bar N_c$}) and in the ({\bf$\bar N_c, N_c$}) representations of the gauge group \cite{kw}. The matter fields form two $SU(2)$ doublets and interact through a quartic superpotential
\be
W_{KW} = \epsilon^{ij}\epsilon^{kl} [A_iB_kA_jB_l]\,.
\ee
Here and in the following, traces over color indices are implied.
In the $N_c=1$ case it is not difficult to show that the moduli space of the theory is in fact described by a conifold singularity, with the following map between geometrical data and mesonic vevs
\be
z_1 = A_1 B_1, \quad z_2 = A_2 B_2, \quad z_3 = A_1B_2, \quad z_4 = A_2B_1 \,.
\label{maps}
\ee
The conifold is a 6d Calabi-Yau cone over the $T^{1,1}$ Sasaki-Einstein manifold. Its Ricci flat metric is usually written as
\bear
ds^2_{C}&=&dr^2+r^2ds^2_{T^{1,1}}\,, \nonumber \\
ds^2_{T^{1,1}}& =& {1\over6}\sum_{i=1}^2[d\theta_i^2+ \sin^2\theta_id\varphi_i^2] + {1\over9}[d\psi+\sum_{i=1}^2\cos\theta_id\varphi_i]^2\,,
\eear
where the range of the angles is $\psi \in [0,4\pi)$,
$\varphi_i \in [0,2\pi)$, $\theta_i \in [0,\pi]$.
In terms of these coordinates we can write
\bear
z_1 = r^\frac32 e^{\frac{i}{2}(\psi - \varphi_1 - \varphi_2)}
\sin\frac{\theta_1}{2}\sin\frac{\theta_2}{2}\,\,,\qquad
z_2 = r^\frac32 e^{\frac{i}{2}(\psi + \varphi_1 + \varphi_2)}
\cos\frac{\theta_1}{2}\cos\frac{\theta_2}{2}\,\,,\rc
z_3 = r^\frac32 e^{\frac{i}{2}(\psi + \varphi_1 - \varphi_2)}
\cos\frac{\theta_1}{2}\sin\frac{\theta_2}{2}\,\,,\qquad
z_4 = r^\frac32 e^{\frac{i}{2}(\psi - \varphi_1 + \varphi_2)}
\sin\frac{\theta_1}{2}\cos\frac{\theta_2}{2}\,\,.
\label{zetas}
\eear
For our purposes of generalizing the background to an unquenched setup,
it is more convenient to work with a redefined radial coordinate
\be
r=r_0 e^\rho\,,
\label{r0def}
\ee
where $\rho$ is a dimensionless quantity and $r_0$ is a constant
which, in order to simplify notation, we set to one from now on. The supergravity background sourced by the $N_c$ D3 branes has a metric of the known warped form
\be
ds^2 = h^{-1/2}(\r)[dx_\mu dx^\mu] + h^{1/2}(\r)
[e^{2\rho}d\r^2 + e^{2\rho} ds^2_{T^{1,1}}]\,, 
\label{kwmetric}
\ee
constant dilaton $\phi$, and $N_c$ units of $F_5$ RR flux through $T^{1,1}$. In the decoupling limit, where the geometry is dual to the Klebanov-Witten fixed point, the relevant metric is $AdS_5\times T^{1,1}$.

The D7-branes used to flavor the KW theory, are taken to be extended along the Minkowski 4d directions and along a non compact four-dimensional submanifold in the transverse space. This makes relatively easy the task of solving for the embedding equations. In fact the warp factor drops out in the D7 DBI action and the spacetime effectively seen by the brane is a direct product of Minkowski spacetime and the conifold. This allows to write the embedding equations in a relatively simple way in terms of the conifold coordinates $z_i$.  A detailed study of the holomorphic D7 embeddings on the conifold can be found in \cite{Ouyang:2003df,kuper,Arean:2004mm,levi}. 

Two classes of embeddings, having $z_1=\mu$ and $z_1-z_2=\mu$ as representative elements, were mainly considered in the literature. On the field theory side the two classes correspond to the addition, to the KW superpotential, of extra cubic and quartic terms, respectively, as well as of mass terms for the fundamental flavors when $\mu\neq0$. In the following we will focus on the first class only. In Appendix \ref{apporb} we will briefly review the field theory interpretation of both kind of embeddings, starting from an orbifold construction.

\subsection{Massless embeddings}
\label{sec: massless}
The simplest embedding we want to consider is $z_1=0$.
We call this embedding massless since it is extended down
to $\r=-\infty$.
From (\ref{zetas}) above, it obviously implies
$\theta_1=0$ or $\theta_2 = 0$. Thus, there are two branches.
In each branch, the D7 brane fully wraps one of the two two-spheres of $T^{1,1}$ and chooses a point in the other one. Notice that usually choosing a point in a $S^2$ requires two equations (fixing $\theta, \varphi$), but of course if one is at a pole ($\sin\theta = 0$), it is not necessary to specify $\varphi$. Thus we can describe a massless embedding of this
kind by choosing two points, one for each sphere, fixing the position
of each branch.

From the field theory point of view the two branches correspond to two classes of flavor multiplets, which we denote by $\tilde q_1, q_1$ and $\tilde q_2, q_2$. The addition to the Klebanov-Witten theory of
$N_f$ D7-branes with $z_1=0$ embedding, modifies the field theory superpotential as \cite{Ouyang:2003df}
\be
W_{z_1=0} = W_{KW} + h_1{\tilde q_1}A_1q_2 + h_2 {\tilde q_2} B_1 q_1\,,
\ee
where, here and in the following, sums over the $N_f$ flavor indices are implied. The classical flavor symmetry preserved by $W_{z_1=0}$ is $U(N_f)\times U(N_f)$.
Because of the $SU(2) \times SU(2)$ symmetry, one should be able to
pick a (two-branch) embedding by picking any point on each of the spheres.
This means that we should have a four-parameter family of massless embeddings.
This family is described by the more general embedding equation \cite{Ouyang:2003df}
\be
\sum_{i=1}^4 \alpha_i z_i = 0\,, 
\label{general_massless}
\ee
with the complex constants $\alpha_i$ spanning a conifold (up to overall complex rescalings)
\be
\alpha_1 \alpha_2 - \alpha_3 \alpha_4 =0\,.
\label{ouyang_eq}
\ee
Notice that embeddings like $z_1-z_2=0$ are not in this family.\footnote{A $SU(2)\times SU(2)$ rotation, in fact, maps $z_1-z_2=0$ into a generalized embedding equation as in (3.9), with the
parameters spanning a unit 3-sphere: $\alpha_1 = -\bar\alpha_2$, $\alpha_3=\bar\alpha_4$, $|\alpha_1|^2+|\alpha_3|^2=1$.} In (\ref{general_massless}), (\ref{ouyang_eq}), we can rescale the coefficients to fix\footnote{Obviously this does not work for the
$\alpha_1 = 0$ case, which implies that one of the branches
is at the south pole of one sphere. This is a zero measure
subset of the embeddings considered.} $\alpha_1=1$. 
Then (\ref{ouyang_eq}) fixes $\alpha_2$.
Thus, this class of embeddings is parameterized by two complex
numbers $\alpha_3, \alpha_4$, as we expect in order to be able
to choose a point in each of the spheres for each branch.
 Indeed, the embedding equation
\be
 z_1 + \alpha_3 \alpha_4 z_2 + \alpha_3 z_3 + \alpha_4 z_4 =0\,,
\label{genmless}
\ee
nicely factorizes into two branches
\be
\left( \sin\frac{\theta_1}{2}
+ \alpha_3 e^{i \varphi_1} \cos \frac{\theta_1}{2}\right)
\left( \sin\frac{\theta_2}{2}
+ \alpha_4 e^{i \varphi_2} \cos \frac{\theta_2}{2}\right)=0\,,
\label{genmlessex}
\ee
so, as expected, $\alpha_3$, $\alpha_4$ determine at which 
point each branch is sitting in each sphere:
\bear
\theta_1 = \theta_1^\infty \equiv 2 \arctan |\alpha_3|\,,\qquad
\varphi_1 = \varphi_1^\infty \equiv \pi - \arg[\alpha_3]\,,\rc
\theta_2 =  \theta_2^\infty \equiv 2 \arctan |\alpha_4|\,,\qquad
\varphi_2 = \varphi_2^\infty \equiv \pi - \arg[\alpha_4]\,.
\label{alphasinf}
\eear
The constants $\theta_i^\infty,\varphi_i^\infty$ denote the position
of each branch as $\rho \to \infty$ (a notation that will be useful
in the next section). We can rewrite (\ref{genmlessex}) as
\be
\Gamma_1 \Gamma_2 =0\, ,
\label{genmless2}
\ee
where we have defined
\be
\Gamma_i \equiv
 \cos\frac{\theta_i^\infty}{2}\sin\frac{\theta_i}{2}
- e^{i (\varphi_i - \varphi_i^\infty)}\sin \frac{\theta_i^\infty}{2} 
 \cos \frac{\theta_i}{2}\,,\qquad (i=1,2)\, .
\label{Gammaidef}
\ee
For later reference, it is worth remarking that
\be
 0 \leq |\Gamma_i| \leq 1\, .
\label{omegalims}
\ee

\subsection{Massive embeddings}

We now want to give a mass to the flavors. This deforms the embeddings
in such a way that the two branches of the massless embeddings merge (the $U(N_f) \times U(N_f)$ flavor symmetry
is explicitly broken down to $U(N_f)$). Let us again start 
with the simplest embedding $z_1 = e^{\frac32\rho_q}
\,e^{i\beta}$, where $\rho_q, \beta$ are real numbers. 
This is referred to as the unit winding ($n_1=n_2=1$) 
embedding in the notations of \cite{Arean:2004mm}
and it is explicitly given by
\be
\psi = \varphi_1 + \varphi_2 + 2\beta\,\,,\qquad
e^{3\rho} = \frac{e^{3\rho_q}}{\left( \sin\frac{\theta_1}{2} \right)^2
\left( \sin\frac{\theta_2}{2} \right)^2} \,\,.
\label{massive1}
\ee
Clearly, $\rho_q$ is the minimal value of $\rho$ reached by the brane.
This is a connected embedding which for large
$\r$ goes to the two branches of the corresponding massless embedding.

We now want to consider the same generalized class for the 
massive embeddings as we used in section
\ref{sec: massless}. The defining equation is \cite{Ouyang:2003df}
\be
z_1 + \alpha_3 \alpha_4 z_2 + \alpha_3 z_3 + \alpha_4 z_4 = {\rm const}\,.
\label{general_massive}
\ee
In terms of the angles, we can write it as
\be
e^{\frac32\rho} e^{\frac{i}{2}(\psi - \varphi_1 - \varphi_2)}
\Gamma_1 \, \Gamma_2=
e^{\frac32\rho_q}
\,e^{i\beta}\,\,,
\label{massive_emb_angles3}
\ee
with the same $\Gamma$'s defined in (\ref{Gammaidef}).
Something obvious that is worth noting is that the embedding
equation is not independent on $\psi$ any more, as expected 
because a mass term breaks the $U(1)_R$ symmetry.
Shifting $\psi$ by a constant shifts the phase $\beta$.
Consistently, there is a $\frac12$ in front of $\psi$ such that
a $4\pi$ shift which takes us to the same point, shifts the mass phase
by $2\pi$, thus leaving it invariant.

Thus, the embedding of a brane within this family depends on 
6 real parameters: $\theta_1^\infty$, $\varphi_1^\infty$,
$\theta_2^\infty$, $\varphi_2^\infty$, $\beta$, $\rho_q$.
In the field theory, the first four should be identified with
the couplings in the superpotential while the last two should
be related to the phase and the modulus of the mass term.

The generalized embeddings we have found above, can be mapped to a field theory superpotential of the form
\be
W = W_{KW} + h_1{\tilde q_1}[A_1+\alpha_4 A_2]q_2 + h_2{\tilde q_2}[B_1 + \alpha_3 B_2]q_1  + m_1 {\tilde
q_1} q_1 + m_2 {\tilde q_2}q_2\,\,,
\ee
where we put generic $m_1, m_2$ mass terms (then we
will want $m_1=m_2$). We can formally rewrite $W$ in a compact form as
$W = W_{KW} + \tilde q M q$ where $M$ is the effective mass matrix for the flavors. Following the arguments in \cite{Ouyang:2003df}, the equation $\det M=0$ should correspond to the D7 embedding equation. Indeed, it reads 
\be
h_1 h_2 [A_1+\alpha_4 A_2] [B_1 +\alpha_3 B_2] = m_1 m_2\,,
\ee
which is equivalent (using, for the $N_c=1$ case, the maps (\ref{maps})) to the equation
\be
 z_1 +\alpha_3\alpha_4 z_2 + \alpha_3 z_3 +\alpha_4 z_4 = {m_1m_2\over h_1h_2}\,\,.
\ee
This is exactly of the form we proposed above, for the massless $m_1=m_2=0$ (\ref{genmless}) and the massive (\ref{general_massive})
case. 

\section{Smearing the embeddings}
\label{sect: sme}
\setcounter{equation}{0}

We want now to consider $N_f \gg 1$ branes suitably distributed
within the family of embeddings (\ref{massive_emb_angles3}).
In particular, we want to restore effectively the $SU(2)\times SU(2)
\times U(1)$ symmetry of $T^{1,1}$ which is broken by flavor branes on top of each other.

Accordingly, we want to place branes at different values of
$\theta_1^\infty, \varphi_1^\infty,
\theta_2^\infty, \varphi_2^\infty, \beta$, while, in principle,
we are interested in keeping $\rho_q$ fixed (clearly, it is also
possible to  implement
a distribution of $\rho_q$'s).
The symmetry preserving density of branes is
\be
\rho_{\theta_1^\infty\varphi_1^\infty
\theta_2^\infty \varphi_2^\infty \beta}\equiv
\frac{dn}{d\theta_1^\infty d\varphi_1^\infty
d\theta_2^\infty d\varphi_2^\infty d\beta}=
\frac{N_f}{2\pi \ (4\pi)^2}\sin \theta_1^\infty \sin \theta_2^\infty\,\,,
\label{massive_density}
\ee
which is normalized as
\be
\int_0^\pi d\theta_1^\infty \int_0^{2\pi} d\varphi_1^\infty 
\int_0^\pi d\theta_2^\infty \int_0^{2\pi} d\varphi_2^\infty
 \int_0^{2\pi} d\beta\ \rho_{\theta_1^\infty\varphi_1^\infty
\theta_2^\infty \varphi_2^\infty \beta} = N_f\,\,.
\ee
The purpose of this section is to compute how the charge
density produced by this D7-brane distribution
 modifies the Bianchi identity for $F_1$.
This coupling comes from the WZ
term, which can be written as (cfr. eq (3.1) of \cite{Benini:2006hh})
\be
S_{WZ} = T_7 \sum_{N_f} \int_{{\cal M}_8} \hat C_8\, \rightarrow\, T_7  \int_{{\cal M}_{10}} \Omega\wedge C_8\,\,,
\ee
yielding:
\be
dF_1 = - T_7 (2\kappa_{(10)}^2)	\Omega = - g_s \Omega\,\,.
\label{gen_bianchi}
\ee

\subsection{General procedure}

Let us start by considering a generic distribution of D7-branes. The $\Omega$ defined above is built from the
orthogonal planes to the submanifolds where the branes are
sitting.
Each brane is described by two equations
\be
f_1(p_i;x^\mu)=0\,\,,\qquad
f_2(p_i;x^\mu)=0 \,\,,
\ee
where the $p_i$ are some parameters and $x^\mu$ are the spacetime coordinates ($\mu=0,...,9$). Locally, the orthogonal plane to a single embedding is described by a two-form:
\be
\delta(f_1) \delta(f_2) df_1 \wedge df_2\,\,.
\label{orth}
\ee
We now need to sum over all the branes. This becomes just an integral
over the parameters which define the embeddings, suitably
weighted with the density:
\be
\Omega = \int \rho_{p_i}(p_i) \left(\delta(f_1) \delta(f_2) df_1 \wedge df_2
\right) dp_i\,\,.
\label{generalOmega}
\ee

\subsubsection{A simple example}

To see how the calculation is performed, let us work out the expression
(\ref{generalOmega}) for a case in which we already know the
answer: the smeared massless embedding in the KW case, 
as discussed in \cite{Benini:2006hh}.
There, since each embedding is disconnected into two branches, we
have to compute separately each contribution.

Consider the collection of branches described by
\be
f_1 (\theta_1^\infty, \varphi_1^\infty; x^\mu) = \theta_1 - \theta_1^\infty\,\,,
\qquad
f_2 (\theta_1^\infty, \varphi_1^\infty; x^\mu) = \varphi_1 - \varphi_1^\infty\,\,.
\ee
The distribution density is
\be
\rho_{\theta_1^\infty\varphi_1^\infty} = \frac{N_f}{4\pi} \sin\theta_1^\infty\,\,.
\ee
We plug this in (\ref{generalOmega}) to get
\be
\Omega = \left(\int \frac{N_f}{4\pi} \sin\theta_1^\infty
\delta(\theta_1 - \theta_1^\infty) \delta(\varphi_1 - \varphi_1^\infty)
d\theta_1^\infty d\varphi_1^\infty\right) d\theta_1 \wedge d\varphi_1=
\frac{N_f}{4\pi}\sin\theta_1 d\theta_1 \wedge d\varphi_1\,.
\ee
From the other branch, we get a similar contribution with ($1\to 2$),
so $\Omega= \frac{N_f}{4\pi}(\sin\theta_1 d\theta_1 \wedge d\varphi_1
+ \sin\theta_2 d\theta_2 \wedge d\varphi_2)$.
This, together with (\ref{gen_bianchi}) above, agrees with equation (2.13)
of \cite{Benini:2006hh}.

\subsection{The Bianchi identity for the smeared massive case}

We now want to compute the expression (\ref{generalOmega})
where $f_1,f_2$ can be read from (\ref{massive_emb_angles3}) and 
the density is written in (\ref{massive_density}). Let us split the modulus and the phase in (\ref{massive_emb_angles3}). From equating the phases, we get
\be
f_1 = \psi - \varphi_1 - \varphi_2 + 2\arg(\Gamma_1)
+ 2\arg(\Gamma_2) - 2\beta -4 \pi n\,\,,
\ee
where $n$ is any integer number. 

From the modulus (squared) we find
\be
f_2 = e^{3\rho} |\Gamma_1|^2 |\Gamma_2|^2-e^{3\rho_q}\,\,,
\label{f2def}
\ee
where
\be
|\Gamma_i|^2=\frac12\left[1-\cos \theta_i^\infty\cos \theta_i
 -  \sin \theta_i \sin \theta_i^\infty
\cos(\varphi_i - \varphi_i^\infty) \right]\,\,.
\label{Gammamodulus}
\ee
We now have to insert these expressions, together with (\ref{massive_density}),
 into (\ref{generalOmega}).
Let us start by making
the integral in $\beta$. Since the only dependence from 
$\beta$ in the integrand comes from $\delta(f_1)$, we have
$\int \delta(f_1) d\beta = \frac12$ (notice that $f_1$ changes
by $4\pi$ when $\beta$ goes from $0$ to $2\pi$, so, for any
values of $\psi, \theta_i, \varphi_i, \theta_i^\infty, \varphi_i^\infty$,
 the argument of the delta function is zero exactly one time in the
$\beta$ range of integration). Thus:
\be
\Omega = \frac{N_f}{(4\pi)^3} \int \left(\sin\theta_1^\infty
\sin\theta_2^\infty\delta(f_2) df_1 \wedge df_2\right) 
d\theta_1^\infty d\varphi_1^\infty d\theta_2^\infty d\varphi_2^\infty\,\,.
\label{Omegamassive}
\ee
Notice that, because of (\ref{omegalims}), $f_2$ can only vanish if
$\rho > \rho_q$. Therefore,  $\Omega (\rho < \rho_q) = 0$, as
expected, since the flavor branes are extended in $\rho > \rho_q$ only. 

The integral (\ref{Omegamassive})
seems extremely difficult to compute. However, 
the result must be fairly simple, just of the form
displayed in (3.99) of \cite{Benini:2006hh} for a 
certain\footnote{Notice that $N_f$ will denote the fixed number of flavor branes while $N_f(\rho)$
is a non-trivial function, representing the effective number of massless flavors
at a given energy scale. These two quantities should not be confused, they
are only equal in the massless case, $\rho_q = -\infty$.} $N_f(\r)$ (a dot will denote derivative with respect to $\rho$ throughout
the paper):
\be
\Omega = \frac{N_f(\r)}{4\pi}(\sin\theta_1 d\theta_1 \wedge d\varphi_1
+ \sin\theta_2 d\theta_2 \wedge d\varphi_2)
-\frac{\dot N_f(\r)}{4\pi} d\r\wedge(d\psi + \cos \theta_1 d\varphi_1 + 
\cos \theta_2 d\varphi_2)\,.
\label{massive_2form}
\ee
 This is the only possibility for an exact two-form preserving
$SU(2) \times SU(2) \times U(1)_\psi \times \IZ_2$ (the $\IZ_2$ interchanging
the $1$ and $2$ $S^2$'s).
This observation will allow us to obtain an exact simple expression
for $\Omega$ since this task is just reduced to computing
a single function $N_f(\r)$.

\begin{figure}[t]
\begin{center}
\includegraphics[width=.4\textwidth]{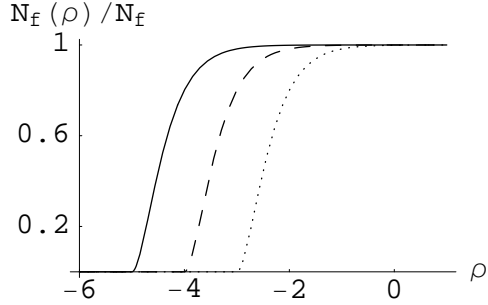}
\end{center}
 \caption{The function $N_f(\r)$ for three different values of the mass parameter (resp. from left to right) $\rho_q=-5\,,-4\,,-3$.}
\label{nfKW}
\end{figure}

Let us expand $\Omega = \frac12\Omega_{MN} dx^M \wedge dx^N,$
where the $M,N$ are the coordinates $\r,\psi,\theta_1,\theta_2,
\varphi_1,\varphi_2$. We can easily isolate the different terms
of this expression and compute them in turn. The
simplest component is
\be
\Omega_{\r\psi} = -\frac{N_f}{(4\pi)^3} \int \left(\sin\theta_1^\infty
\sin\theta_2^\infty \delta(f_2) 3 e^{3\rho} |\Gamma_1|^2|\Gamma_2|^2\right)
d\theta_1^\infty d\varphi_1^\infty d\theta_2^\infty d\varphi_2^\infty\,\,.
\ee
Still, this is not so easy to compute due to the delta function. 
But, since $ \Omega_{\r\psi}$ cannot depend on $\theta_i, \varphi_i$
(see (\ref{massive_2form})), we can fix $\theta_1=\theta_2=0$ such
that, from (\ref{Gammamodulus}), we have
$|\Gamma_i|^2 = \frac12 (1- \cos\theta_i^\infty)$.
Changing variables to $a_i = \frac12 (1-\cos \theta_i^{\infty})$
(notice that $0\leq a_i \leq 1$), the integral above now takes a simple form:
\be
\Omega_{\r\psi} = -\frac{N_f}{4\pi}
\int \delta(e^{3\r}a_1 a_2 - e^{3\r_q})
3e^{3\r}\,a_1\, a_2\, da_1\, da_2 =
- \frac{3N_f}{4\pi} e^{3\r_q-3\r} \int_{e^{3\r_q-3\r}}^1
\frac{da_2}{a_2}\, .
\ee
Comparing the result to (\ref{massive_2form}) we get the sought
result:
\bear
\dot N_f(\r) &=& 3 N_f  e^{3\r_q-3\r} (3\rho-3\rho_q)\,\,,\qquad\qquad
\rc
N_f(\r) &=& N_f \left[1 - e^{3\r_q-3\r} ( 1 +3\r - 3\r_q)\right]
\,\,,\qquad
(\rho > \rho_q)
\label{Nfofr}
\eear
whereas $N_f(\r) = 0$ at $\rho < \rho_q$.
Consistently, $N_f(\infty)=N_f$ and $N_f (\rho_q)=0$, so
$N_f(\r)$ is a continuous function which asymptotes to
the one of the massless case when $\r \gg \rho_q$. 
Notice also that $N_f(\r), \dot N_f(\r)\geq 0 $
as required on intuitive grounds and also by the 
supergravity  equations of motion \cite{Benini:2006hh}. 
As it is shown in figure \ref{nfKW}, 
$N_f(\rho)$ has the shape of a smoothed-out Heaviside step function.

This completes the computation of the $\Omega$ since the symmetry
constrains the rest of the components to be given by
(\ref{massive_2form}). We have checked 
the result (\ref{Nfofr}) by calculating explicitly, by
numerical integration, the other components. 

The effective running of the number of flavors has a natural field theory interpretation. At energy scales larger than the mass of the matter multiplets, the theory looks like the massless one. At energies lower than the mass scale, the massive flavors can be integrated out and the theory resembles the unflavored ($N_f=0$) one. 
At scales around $\rho \sim \rho_q$, the non-trivial profile of the function
should take into account threshold effects.

\section{The backreacted solution with massive flavors}
\label{sect: thesol}
\setcounter{equation}{0}

Knowing the precise expression for $N_f(\r)$, we can find the backreacted solution dual to the massive flavored KW theory
using the expressions of \cite{Benini:2006hh}. The ansatz for the Einstein frame metric
\be
ds^2 = h^{-\frac12} dx_{1,3}^2 + h^\frac12\left(
e^{2f} d\rho^2 + \frac{e^{2g}}{6}\sum_{i=1,2}
(d\theta_i + \sin^2 \theta_i d\varphi_i^2)
+\frac{e^{2f}}{9}(d\psi + \sum_{i=1,2} \cos \theta_i d\varphi_i)^2\right),
\label{defkwa}
\ee
is a deformation, driven by the functions $f(\rho), g(\rho)$, of the warped conifold considered up to now (for which $f(\rho)=g(\r)=\r$). 
The radial coordinate $\rho$ varies in $(-\infty,\rho_L]$ with the lower (resp. upper) extremum being mapped, via the holographic radius/energy relation, to the extreme IR (resp. UV Landau pole) of the dual field theory.  
The ansatz for the RR-forms and the dilaton reads
\bear
\phi&=&\phi(\rho)\,\,, \nonumber \\
F_5 &=& 27 \pi g_s N_c \a'^2 e^{-4g-f} \, h(\rho)^{-5/4} \Big( e^{x^0x^1x^2x^3\rho} -
e^{\theta_1\varphi_1\theta_2\varphi_2\psi}
\Big)\,\,, \nonumber \\
F_1 &=& \frac{g_sN_f(\rho)}{4\pi} \bigl( d\psi + \cos\theta_1\,
d\varphi_1 + \cos\theta_2\, d\varphi_2 \bigr)\,\,, \nonumber  \\
dF_1 &=& -\frac{g_sN_f(\rho)}{4\pi}(\sin\theta_1 \, d\theta_1\wedge d\varphi_1 +
\sin\theta_2 \, d\theta_2\wedge d\varphi_2 \bigr) + \nonumber \\ 
&& +\frac{g_s\dot N_f(\rho)}{4\pi} d\rho\wedge(d\psi + \cos \theta_1 d\varphi_1 + 
\cos \theta_2 d\varphi_2)\,\,.
\eear
The vielbein is given by
\bear
e^{x^i} &=& h^{-1/4} \, dx^i, \quad  e^{\rho} = h^{1/4}e^f d\rho\,\,, \nonumber \\
e^{\theta_i} &=& \frac{1}{\sqrt{6}} h^{1/4} e^g d\theta_i, \quad  e^{\varphi_i}= \frac{1}{\sqrt{6}} h^{1/4} e^g \, \sin\theta_i d\varphi_i\,\,, \nonumber \\
e^{\psi} &=& \frac{1}{3} h^{1/4} e^f
( d\psi + \cos\theta_1\, d\varphi_1 + \cos\theta_2\, d\varphi_2)\,\,.
\eear
It is crucial to notice that the projections that define the Killing spinors have the same
form for the deformed or undeformed ansatze \cite{Benini:2006hh}. Then, it is
not difficult to check that the $\kappa$-symmetry analysis of
\cite{Arean:2004mm} is easily generalized to the backreacted solution,
and (\ref{massive1}) is obtained
 without modification
(when one uses the $\rho$-coordinate). 
This means that the computation of section \ref{sect: sme}
and in particular the result (\ref{Nfofr})
is directly applicable to the backreacted case.

The other crucial observation is that passing from $N_f$ to $N_f(\rho)$ does not alter the form of the first order equations for the functions appearing in the ansatz. This is due to the fact that the supersymmetric fermionic variations only contain the forms $F$ and not the $dF$ terms. Thus, although the Bianchi identity for $F_1$ is modified in this massive setup w.r.t. the massless one, we can just use the first order equations found in \cite{Benini:2006hh} with $N_f\rightarrow N_f(\rho)$. 
We thus have:
\bear
\dot g &=& e^{2f-2g}\,\,,\rc
\dot f &=& 3 - 2 e^{2f-2g} - \frac{3 g_s N_f(\rho)}{8\pi} e^\phi\,\,,\rc
\dot \phi &=& \frac{3 g_s N_f(\rho)}{4\pi} e^\phi\,\,,\rc
\dot h &=& - 27 \pi g_s N_c \a'^2\, e^{-4g}\,\,.
\label{sue}
\eear
These equations have to be solved separately for $\rho < \rho_q$
and $\rho > \rho_q$. Then, one has to demand continuity at $\rho = \rho_q$.
This will be the content of the upcoming subsections.

The constituent mass $m_q$ for the dynamical quarks can be related to $\rho_q$ by defining it, as usual, as the energy of a straight string stretched along the radial direction from $\rho_q$ to the bottom of the space, that is at $\rho=-\infty$ in our case:
\be
m_q = {1\over 2\pi\alpha'}\int_{-\infty}^{\rho_q}e^{{\phi\over2}+f}d\rho\,\,.
\label{qm}
\ee

\subsection{Region 1: $\rho < \rho_q$}

In this region, one just has the well known unflavored system,
after inserting $N_f(\r)=0$ in (\ref{sue}). The general solution for
$\phi,g,f$ can be given in a simple form:
\bear
e^\phi &=& e^{\phi_{IR}}\,\,,\rc
e^g &=& \tilde c_3 \left( e^{6\rho} + \tilde c_1 \right)^\frac16 \,\,,\rc
e^f &=& \tilde c_3 e^{3\rho}\left( e^{6\rho} + \tilde c_1 
\right)^{-\frac13}\,\,,\rc
h(\rho)&=& 27 \pi g_s N_c \a'^2(\tilde c_2 +   \int_\rho^{\rho_q} e^{-4g(\rho_*)} d\rho_*)\,\,.
\label{KWunflavoredsol2}
\eear
For the integration constants, we have used a notation similar
to \cite{Benini:2006hh}.

\subsection{Region 2: $\rho > \rho_q$}

The relevant system of equations for this region comes from inserting
$N_f(\rho)$ as given in (\ref{Nfofr}) in (\ref{sue}).
Remarkably, it turns out that the functions $f,g,\phi$ can be explicitly
integrated in this region too. Just as in the massless case, the dilaton is running and blows up at a certain $\rho_L$.
Fixing one of the constants of integration, which just amounts to shifting
$\rho$, $\rho_q$, we require, as in \cite{Benini:2006hh}, that the
dilaton divergence ({\it i.e.} the field theory UV Landau pole) is located at $\rho_L = 0$.
The solution is:
\bear
e^\phi &=& -\frac{4\pi}{g_s N_f\left(3\rho 
- e^{3\rho_q - 3\rho}(3\rho_q -3\rho -2)
+ e^{3\rho_q} (3\rho_q -2)
\right)}\,\,,\rc
e^f&=&c_3 \frac{\left(
-6 e^{6\rho} \rho + 2 e^{3\rho + 3\rho_q}(3\rho_q -3\rho -2)
-2 e^{6\rho + 3\rho_q} (3\rho_q -2)
\right)^\frac12}{\left(
 e^{6\rho} (1-6\rho) +  4 e^{3\rho + 3\rho_q}
(3\rho_q -3\rho -1) - 2 e^{6\rho + 3\rho_q} (3\rho_q -2)+c_1
\right)^\frac13}\,\,,\rc
e^g &=& c_3\left(
 e^{6\rho} (1-6\rho) +  4 e^{3\rho + 3\rho_q}
(3\rho_q -3\rho -1) - 2 e^{6\rho + 3\rho_q} (3\rho_q -2)+c_1
\right)^\frac16\,,\rc
h(\rho)&=& 27 \pi g_s N_c \a'^2(c_2 +   \int_\rho^{0} e^{-4g(\rho_*)} d\rho_*)\,\,.
\label{KWflavoredsoldil}
\eear

\subsection{Patching the solutions at $\rho=\rho_q$}
\label{sec: patching}

Out of the seven integration constants appearing in 
(\ref{KWunflavoredsol2}), (\ref{KWflavoredsoldil}), 
four are fixed by demanding continuity
of the functions at $\rho = \rho_q$, namely:
\bear
e^{\phi_{IR}} &=& \frac{4\pi}{g_s N_f \left(e^{3\rho_q} (2-3\rho_q)-(2+3\rho_q)\right)}\,\,,\rc
c_1 &=& - e^{6\rho_q} + \tilde c_1
\left( 2\ e^{3\rho_q} (2 -3\rho_q) - 2 (2+3\rho_q)\right)\,\,,\rc
\tilde c_3 &=& c_3 
\left( 2\ e^{3\rho_q} (2 -3\rho_q) - 2 (2+3\rho_q)\right)^{\frac16}\,\,,\rc
\tilde c_2 &=& c_2 + \int_{\rho_q}^0 e^{-4g(\rho_*)} d\rho_*\,\,.
\label{matchingconsts}
\eear
Notice that the infrared dilaton blows up, signaling the breaking of the validity of the supergravity approximation, for $\rho_q\rightarrow0$, i.e. for dynamical masses
close to the Landau pole.

\subsection{A prescription for the integration constants}
\label{sec: proposal}

Apart from depending on the parameters $N_c,N_f,\rho_q$, which
have a clear field theory interpretation ($\rho_q$ being related to
the modulus of the quark masses), the solution described above also
contains
three independent integration constants $\tilde c_1$, $c_2$, $c_3$.
In this subsection we will discuss their physical meaning and give
a prescription to fix them when comparing different solutions, as 
we will do in section \ref{sec: Wilson}.

Let us start with $\tilde c_1$, which severely affects the IR behaviors.
If $\tilde c_1=0$ one recovers in the IR limit
the standard regular solution for D3-branes on the conifold. 
If $\tilde c_1>0$ the solution is of the form discussed in \cite{leo}: in this case the transverse space to the D3-branes is a deformation of the ($\IZ_2$ orbifold of the) conifold where a 4-cycle has blown up. The D3's are actually smeared along that cycle and their backreaction gives rise to a singular 10d background. Here, we will
not be interested in such solutions and thus, by demanding IR regularity,
we impose
\be
\tilde c_1 = 0\,\,, 
\ee
which, from (\ref{matchingconsts}), sets
\be
c_1=-e^{6\r_q}\, .
\ee
The constant $c_2$ affects mainly the UV and corresponds to turning on a source term for an irrelevant operator. We will require our solutions 
to share a common UV behavior by setting
\be
c_2 =0 \,,
\ee
such that $h$ precisely vanishes at the Landau pole.
We then have $\phi|_{\r=0}=\infty$, $h|_{\r=0}=e^f|_{\r=0}=0$.

Finally, $c_3$, which simultaneously rescales $e^{2g}, e^{2f}, h^{-\frac12}$,
can be reabsorbed by a rescaling of the Minkowski coordinates. 
We will  adopt an {\it ad hoc} prescription and, as before,
require that all solutions behave similarly in the UV by demanding
$e^g|_{\r=0}=1$. This fixes:
\be
c_3 = (1- e^{6\r_q} + 6 \r_q e^{3\r_q})^{-\frac16}\, .
\ee
Inserting this value in (\ref{matchingconsts}) we obtain:
\be
\tilde c_3 = 
\left(\frac{ 2\ e^{3\rho_q} (2 -3\rho_q) - 2 (2+3\rho_q)}
{1- e^{6\r_q} + 6 \r_q e^{3\r_q}}
\right)^{\frac16}\,\,.
\ee
Let us close this section by inserting these prescriptions in
the definition of the constituent quark mass (\ref{qm}):
\be
m_q = \frac{1}{2\pi \alpha'} \sqrt{\frac{8\pi}{g_s N_f}}
\frac{e^{\rho_q}}{(1- e^{6\r_q} + 6 \r_q e^{3\r_q})^\frac16
(2\ e^{3\rho_q} (2 -3\rho_q) - 2 (2+3\rho_q))^\frac13}\, .
\label{mqvalue}
\ee

Notice that $\lim_{\rho_q \to 0} m_q = \infty$.  
Of course we cannot trust the supergravity formulas in this limit since the dilaton is blowing up. The best we can say is that these results suggest that, within the choice of integration constants adopted above, we can explore a wide range of dynamical mass parameters, from $m_q\rightarrow 0$ to some higher $m_q<\Lambda_{UV}$.
\subsection{The Ricci curvature and regime of validity}
\label{ricci}
The validity of the supergravity approximation that we are using, requires the curvature invariants of the string frame version of (\ref{defkwa}) to be small in $\alpha'$ units. Let us focus, in particular, on the Ricci scalar. Using the BPS equations
(\ref{sue}), one obtains
\be
R_S = -2 \frac{3g_s N_f(\rho)}{4\pi}h^{-\frac12} e^{-2g + \frac{\phi}{2}}
\left(7 + 4 \frac{3g_s N_f(\rho)}{4\pi} e^{2g -2f + \phi}+\frac74
e^{2g-2f} \frac{\dot N_f(\rho)}{N_f(\rho)}\right)\,\,.
\ee
Obviously, it vanishes in region 1 ($\rho < \rho_q$) where the near-horizon ($\rho\rightarrow -\infty$) metric is $AdS_5\times T^{1,1}$ and the dilaton is constant. Validity of supergravity in region 1 requires $\lambda_{IR}=g_s N_c e^{\phi_{IR}}$ to be large and $g_se^{\phi_{IR}}$ to be small, as usual. From (\ref{matchingconsts}) we see that these conditions are satisfied, at fixed $\rho_q$, if $N_c\gg N_f\gg 1$. 

\begin{figure}[t]
\begin{center}
\includegraphics[width=.4\textwidth]{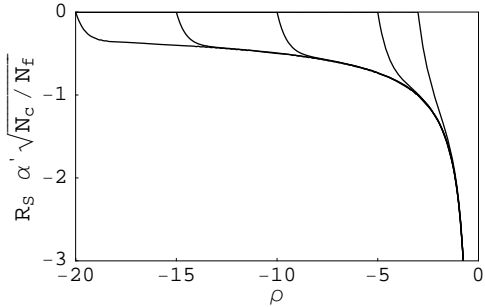}
\end{center}
 \caption{The curvature scalar as a function of $\rho$ for,
from left to right, \mbox{
$\rho_q =-20,-15,-10,-5,-3$}.}
\label{riccifig}
\end{figure}

Let us now concentrate on region 2. Since $e^\phi \sim (g_s N_f)^{-1}$ and $h \sim g_s N_c \a'^2$, it is easy to see that one can write $R_S$ as $\sqrt{N_f/N_c}\, \a'^{-1}$ times a factor which only depends on $\rho$, $\rho_q$ (once the integration constants $c_1,c_2$ have been fixed, notice that $R_S$
does not depend on $c_3$).
Figure \ref{riccifig} depicts some examples, where the
prescription of section \ref{sec: proposal} for $c_1,c_2$ has been used.

In region 2, in order for the supergravity approximation to be valid
in a large range of $\rho$, we
need $N_f \ll N_c$ so that the Ricci scalar stays 
parametrically small except, eventually, very
near the Landau pole.\footnote{This restriction is typical in D3D7 systems \cite{D3D7localized}. Notice that, despite the limit $N_f\ll N_c$, we are not in the quenched approximation, since the backreaction of the flavor branes is taken into account.} Notice also that the only singularity of the solution
is located at the Landau pole 
(a pathological behavior expected on physical grounds). As stressed in
the introduction, the smearing of massive flavors allows to smooth out IR
singularities.

Apart from checking the validity of the supergravity approximation which
requires $N_f \ll N_c$ and $g_s \ll 1$, we should also check the validity
of the smeared approximation and the DBI action, along the lines of
Section \ref{sect: gensme}. The branes we are smearing have codimension
$d=2$ and, as we have seen, the typical length of the transverse space is 
governed by $R\sim (N_c / N_f)^\frac14$. Within an area of $\alpha'$ size
there are $N_f / R^2\approx N_f^\frac32 N_c^{-\frac12}$ flavor branes, a number which should
be larger than 1 for the smearing to be a good approximation
down to $\alpha'$ scales.
On the other hand, for the DBI to be valid, 
$g_s N_f^\frac32 N_c^{-\frac12}\ll 1$. If we take $g_s \sim N_c^{-1}$ or
$N_f^{-1}$,
it is clear that this condition does not impose any further restriction.
We thus get the following range of parameters:
\be
1\ll N_c^\frac13 \ll N_f \ll N_c\, .
\ee
Similar restrictions on the parameters should be considered also in the the massless case $\rho_q=-\infty$. It should be interesting to investigate
if there are regions of $\rho$ for which our solution can be valid for
more general values of $N_f/N_c$ and to check how the curvature - and, so, the validity range of the supergravity approximation - is affected by the choice of the integration constants.
\subsection{Further comments on the dual field theory}
As we have remarked above, the string background we have constructed is dual to a flavored Klebanov-Witten theory which has gauge group $SU(N_c)\times SU(N_c)$, a Landau pole in the UV and is conformal in the IR. This is reflected by the behavior of the dilaton $e^{\phi}$, which, provided the standard orbifold dictionary can be applied to the present setup, is mapped to the field theory gauge couplings by
\be
{4\pi\over g_1^2}+{4\pi\over g_2^2} = g_s^{-1} e^{-\phi}\,.
\ee
Without loss of generality here we consider the case $g^2_1=g^2_2\equiv 
g^2_{FT}=8\pi g_s e^{\phi}$. The dilaton  is running and going to infinity for $\rho\rightarrow 0$, while it is constant in the region $\rho<\rho_q$. The beta function for $g_{FT}$ can be inferred, holographically, from the dilaton equation
\be
{d\over d\rho} e^{-\phi} = - {3 g_s N_f(\rho)\over 4\pi}\,\,,
\ee
once the precise radius/energy relation is given. In the massless case, the relation $\rho=\log{(\mu/\Lambda_{UV})}$ was used to get the field theory beta function \cite{Benini:2006hh}. In our case this relation should still be valid with a good approximation when the flavors are very light. Notice that this relation suggests that
\be
\rho_q = \log{m_q\over\Lambda_{UV}}\,\,,
\ee
which, looking at eq. (\ref{mqvalue}), is just the leading term in the small mass limit where $\rho_q\rightarrow -\infty$.
Within these approximations, we can write the field theory running coupling as
\be
{8\pi\over g_{FT}^2}\approx -{N_f\over4\pi}\left[3\log{\mu\over\Lambda_{UV}}-\left({m_q\over\mu}\right)^3[3\log{m_q\over\mu}-2]+\left({m_q\over\Lambda_{UV}}\right)^3[3\log{m_q\over\Lambda_{UV}}-2]\right]\,.
\ee
In this expression, the first term reproduces the one-loop result, while the rest 
is a string prediction.\footnote{One has to keep in mind that the string theory calculation of $N_f(\rho)$ is reflecting a particular choice of renormalization scheme in the field theory side.} The power-like factors in the above expression could 
be interpreted in field theory as fractional instantons, or renormalon 
corrections; the interesting feature is the presence of just two such 
terms.
\section{The Heaviside approximation}
\label{sect: step}
\setcounter{equation}{0}

A simple way to model the integrating in/out of the massive flavor degrees of freedom in field theory is to forget about the details of the embedding and model the string dual with a Heaviside step function $N_f(\rho)=N_f\Theta(\rho-\rho_q)$. Of course this approximation is a source of systematic errors, due to the differences between the step function and the true value of $N_f(\rho)$ extracted from the smearing procedure. In the present setup we can measure these errors, comparing (see the following sections) physical observables deduced using both the true and the approximate supergravity solutions. Estimating these errors is of notable importance to have an idea of how well the step function approximation can work in general models like \cite{screening}, where the correct $N_f(\rho)$ may be hard to compute.

We now present the solution using the Heaviside form for $N_f(\rho)$,
using capital symbols for integration constants within this framework.
At $\rho < \rho_q$, we still have
(\ref{KWunflavoredsol2}), whereas for
$\rho > \rho_q$ the solution reads:
\bear
e^\phi &=& - \frac{4\pi}{3g_s N_f \rho}\,\,,\rc
e^g &=& C_3\left((1 - 6 \rho) e^{6\rho} 
+C_1 \right)^\frac16\,\,,\rc
e^f &=& C_3\sqrt{-6\rho} \ e^{3\rho} \left((1 - 6 \rho)
e^{6\rho} +C_1 \right)^{-\frac13}\,\,,\rc
h&=& 27 \pi g_s N_c \a'^2 (C_2 +   \int_\rho^0 e^{-4g(\rho^*)} d\rho^*)\,\,.
\label{KWflavoredsol3}
\eear
As in section \ref{sec: patching},  continuity at $\rho_q$
gives some relations among the different constants:
\be
e^{\phi_{IR}} = - \frac{4\pi}{3g_sN_f \rho_q}\,\,,\qquad
C_1 = -6\rho_q \tilde C_1 - e^{6\r_q}\,\,,\qquad
\tilde C_3 = C_3 (-6 \rho_q)^\frac16\,\,,
\ee
while if we follow again the prescriptions of section 
\ref{sec: proposal}, we get:
\be
\tilde C_1 =C_2 = 0\,,\qquad
C_3 = (1-e^{6\rho_q})^{-\frac16}\,\,.
\ee
Under this prescription, the constituent quark mass
(\ref{qm}) reads:
\be
m_q = \frac{1}{2\pi \alpha'}\, \sqrt{\frac{8\pi}{g_s N_f}}\,
\frac{e^{\rho_q}}{(- 6 \r_q)^\frac13
(1 - e^{6  \rho_q})^\frac16}\, .
\label{mqvaluestep}
\ee
Notice that for $\rho_q\rightarrow -\infty$ the solutions obtained with the actual $N_f(\r)$ approach those written above: this suggests that the Heaviside approximation is quite good in the small $m_q$ limit.

\section{Wilson loops in backreacted D3D7 models}
\label{sec: Wilson}
\setcounter{equation}{0}

In the previous sections we have constructed a string dual of a flavored version of the Klebanov-Witten theory with a large number $N_f$ of dynamical flavors of mass $m_q$. Now we want to study how these sea flavors affect the non perturbative dynamics of the gauge theory. We are going to probe the latter with an external quark-antiquark $\bar Q, Q$ pair with mass $M_Q\gg m_q$. Due to the presence of a Landau pole in the gauge theory, $M_Q$ cannot be taken to be infinite and it must lie below the pole. The idea is to study how the static quark-antiquark potential depends on the sea quark parameters $m_q, N_f$. 
The effects of $N_f$ flavors on the energy/spin relations for mesonic-like bound states in a similar but localized D3D7 system have been studied in \cite{Kirsch:2005uy}.

The $\bar QQ$ bound state is dual to an open string with the extrema lying on a probe D7-brane embedded in such a way that it reaches a minimal distance $\rho_Q\gg \rho_q$  from the bottom of the space ($M_Q = {1\over 2\pi\alpha'}\int_{-\infty}^{\rho_Q}e^{{\phi\over2}+f}d\rho$). 
The string, in turn, bends in the bulk and reaches a minimal radial position $\rho_0$. The Minkowski separation $L$ between the test quarks, as well as the total energy
of the system, depends on $\rho_{0}$. From this we can deduce the $V(L)$ relation, where $V(L)$ is the $\bar Q Q$ potential, i.e. the
total energy to which the total contribution from the quark mass $2M_Q$ has been
subtracted. The open string embedding is chosen as $t=\tau,
y=\sigma, \rho=\rho(y)$ where $y\in [-L/2,L/2]$ is one of the
spatial Minkowski directions. The string worldsheet action reads 
\bear
S= -\frac{1}{2\pi\alpha'}\int dt dy \sqrt{g_{tt}(\rho)[g_{yy}(\rho)
+ (\partial_y \rho)^2 g_{\rho\rho}(\rho)]}\,\,, \nonumber \eear 
where $g_{\mu\nu}$ refers here to the string frame metric, which is
obtained multiplying (\ref{defkwa}) by $e^\frac{\phi}{2}$.
Defining
\be
F = \sqrt{g_{tt}g_{yy}}= e^{{\phi\over2}} h^{-1/2}\,,\quad\quad G = \sqrt{g_{tt}g_{\rho\rho}} = e^{{\phi\over2}} e^f\,,
\ee 
we can write \cite{maldawilson,sonne} the string length and renormalized energy as 
(the $0$ subindex means that
the quantity is evaluated at $\rho=\rho_0$)
 \bear L(\rho_0)&=&2
\int_{\rho_0}^{\rho_{Q}} \frac{G F_{0}}{F\sqrt{F^2
-F_{0}^2}}d\rho\,\,,\rc V (\rho_0)&=&\frac{2}{2\pi\alpha'} \Bigl[\int_{\rho_0}^{\rho_{Q}}
\frac{G F}{\sqrt{F^2 -F_{0}^2}}d\rho -
\int_{0}^{\rho_{Q}} G \ d\rho \Bigr]. \nonumber \eear
We can use the background solutions found in the previous sections to study how the external quark interaction depends on the dynamical massive flavors. 
In doing so, we vary one of the physical parameters $N_f,  m_q, M_Q$ while keeping the other two fixed.
In the numerical computations we will present, lengths will be measured
in units of $\sqrt{27\pi g_s N_c \a'^2}$ whereas energies in units
of $(2\pi \a')^{-1}$ (the $r_0$ that was set to 1 in (\ref{r0def}) should be
reinserted to get the correct dimensions).
\begin{figure}
\centering
\includegraphics[width=0.3\textwidth]{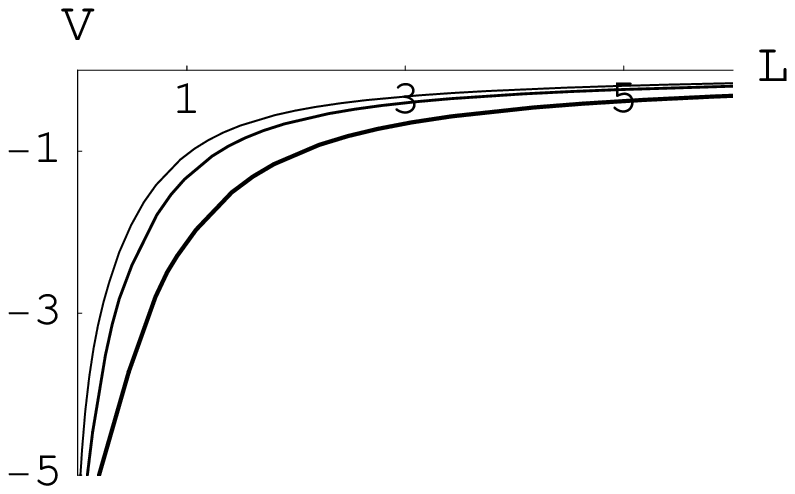}
\includegraphics[width=0.3\textwidth]{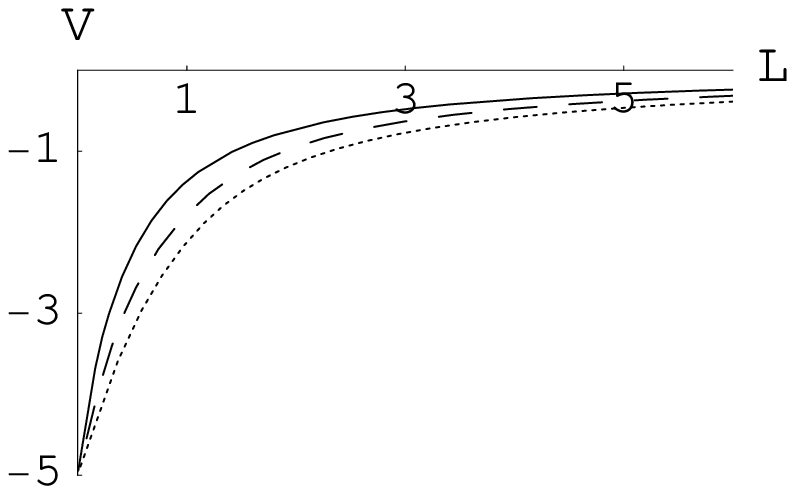}
\includegraphics[width=0.3\textwidth]{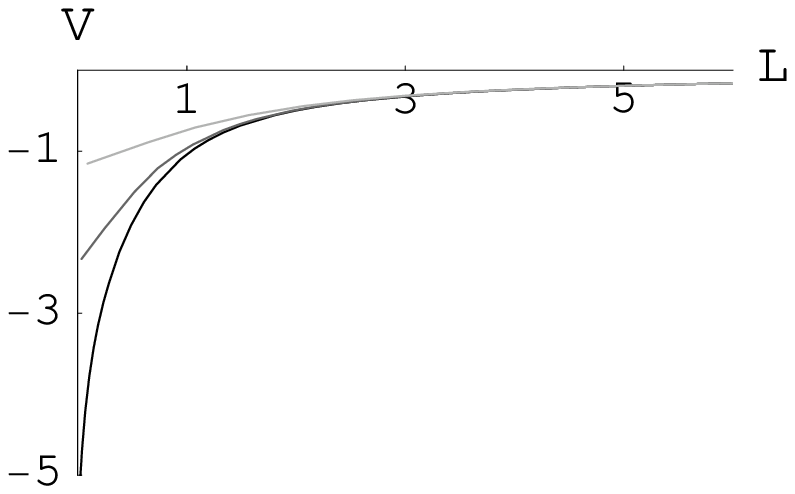}
\caption{From left to right, plots of $V(L)$ at different values of: $g_s
N_f=1, 0.6, 0.2$, top to bottom (at $m_q=0.1$, $M_Q=3$); $m_q=0.5, 1, 1.5$, top to bottom (at $g_s N_f=1$, $M_Q=2.5$); $M_Q=0.6, 1.2, 3$, top to bottom (at $g_sN_f=1$, $m_q=0.1$).}
\label{VLplots}
\end{figure}

From figure \ref{VLplots} we can see that the static quark-antiquark potential is always negative and has the same qualitative Coulomb-like behavior as in the unflavored KW conformal case. This behavior is not unexpected: in the far IR (i.e. at large $L$) the quarks are integrated out and the theory looks like the KW one. Hence at large $L$ we expect the Coulomb-like behavior typical of a conformal theory. Figure \ref{VLplots} also tell us that the absolute value of the potential $|V(L)|$ at fixed $L$ is a decreasing function of $N_f$ and an increasing function of $m_q$. This is an effect of the screening of the color charges due to the dynamical flavors: the more the theory is ``unquenched'' (large $N_f$, small $m_q$) the more the modulus of the quark-antiquark force is reduced. Finally, the large $L$ behavior of $V(L)$ is not strongly affected by the choice of the cutoff $M_Q$; in the small $L$ region, instead, we see that $|V(L)|$ is an increasing function of $M_Q$.
We will see in section \ref{comp} how these results nicely fit with the expected qualitative behavior of $V(L)$.

For all the plots of this section, we have adopted the prescription for
integration constants given in section \ref{sec: proposal}.
It can also be instructive to study how the behavior of the $\bar Q Q$ interaction is affected by the choice of these integration constants. A brief analysis of
the dependence on $c_2$ can be found in Appendix \ref{app: integr}.

\subsection{The screening length}
The connected $\bar Q Q$ configuration we have studied above is expected to be unstable due to the presence of the dynamical flavors. If the $\bar Q Q$ string has enough energy, in fact, a  dynamical $\bar q q$ pair can be popped out from the vacuum with a consequent decay of the heavy $\bar Q Q$ state into a pair of heavy-light mesonic-like bound states $\bar Qq + \bar q Q$.
\begin{figure}[t]
\centering
\includegraphics[width=.3\textwidth]{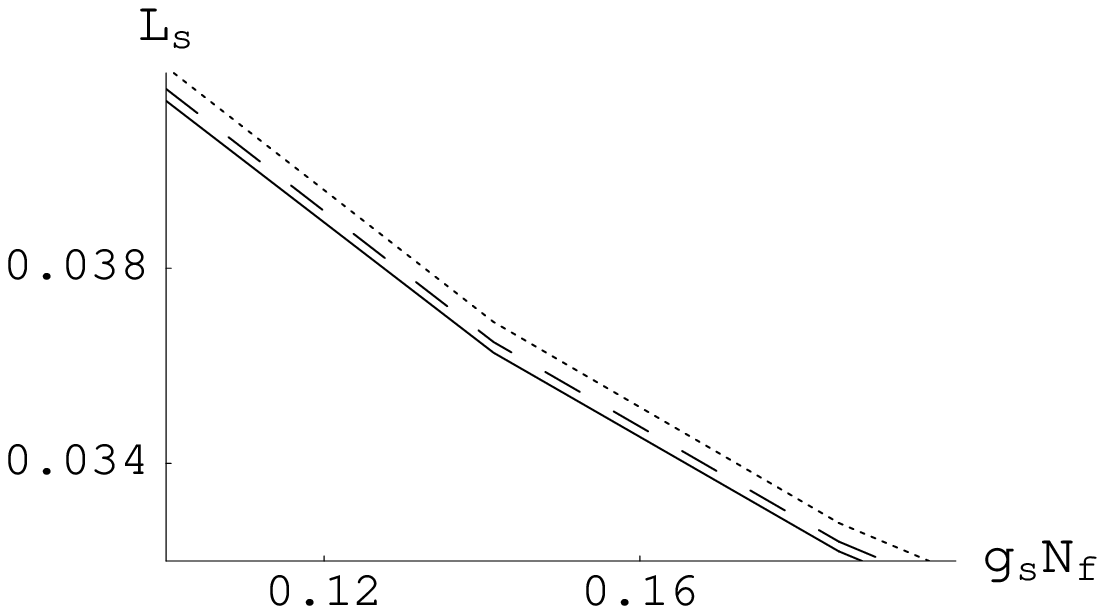} 
\includegraphics[width=.3\textwidth]{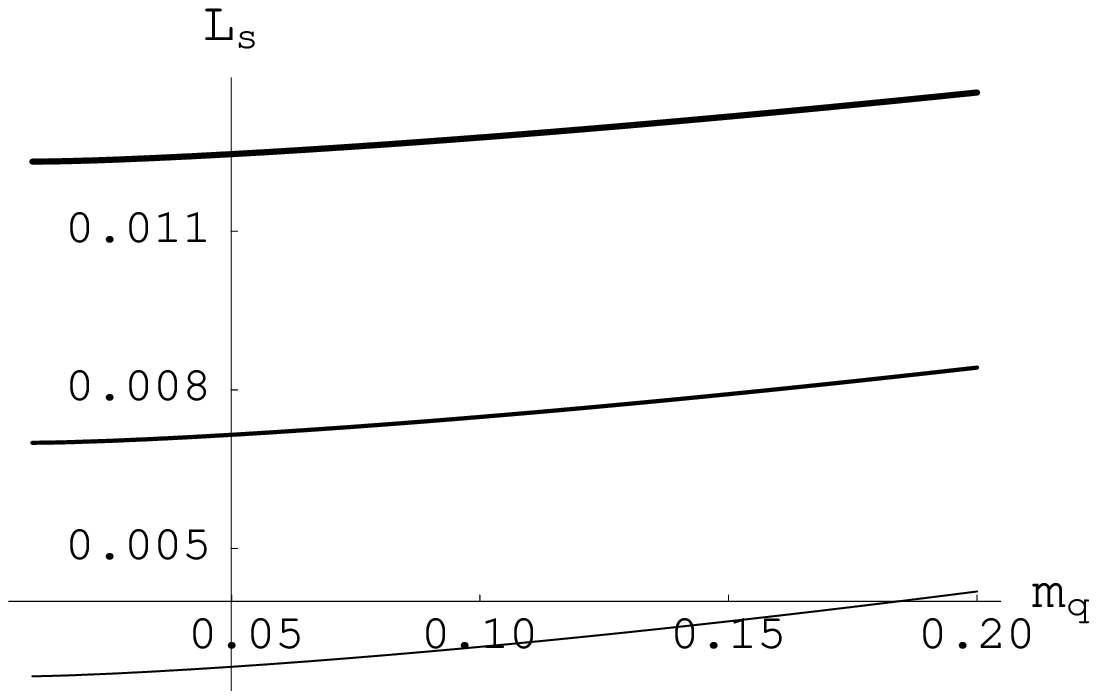}
\includegraphics[width=.3\textwidth]{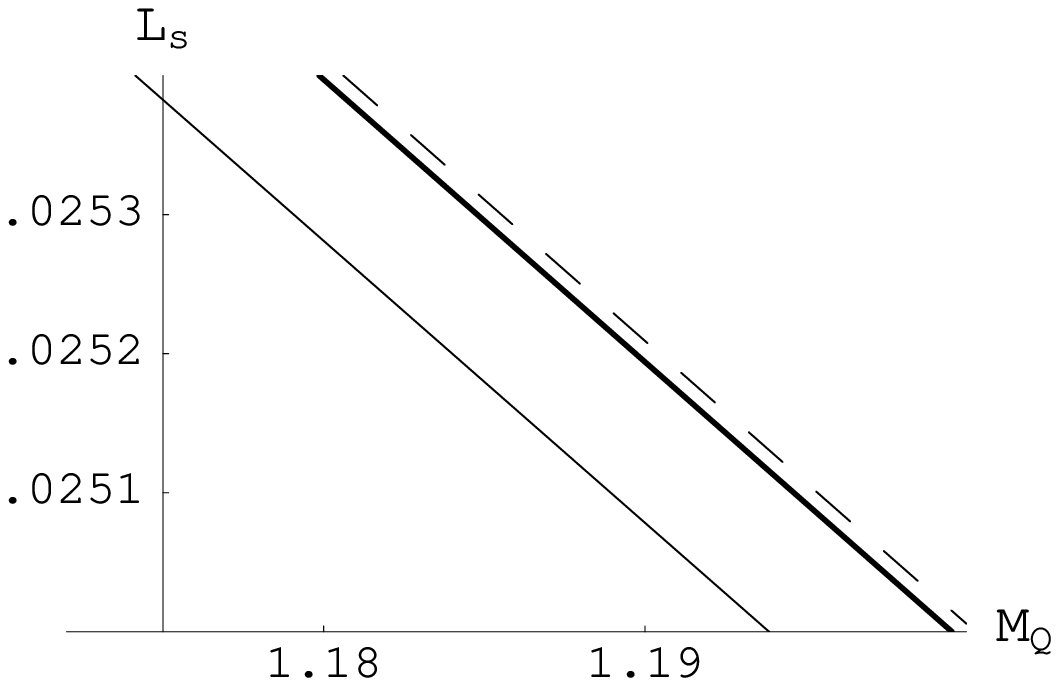}
\caption{From left to right, the screening length $L_{s}$ as a function of: $g_sN_f$ for $m_q=0.01, 0.05, 0.1$, bottom to top ($M_Q=2$); $m_q$ for $g_sN_f=1, 1.5, 2$, top to bottom ($M_Q=2$); $M_Q$ for $g_sN_f=1, m_q=0.1$ (thick), $g_sN_f=1, m_q=0.101$ (dashed), $g_sN_f=1.01, m_q=0.1$ (solid).}
\label{Ls}
\end{figure}
The length at which the decay can happen is the so called ``screening length'' $L_s$. In the model at hand, due to the smearing procedure, the lighter heavy-light ``mesons'' are nearly massless. This is due to the fact that there exist dynamical flavor branes which intersect the probe brane corresponding to the test quarks. The open string stretching between the probe and those dynamical branes has nearly massless modes at the intersection. In fact their mass is expected  \cite{herzog} to scale as $M_Q/\sqrt{\lambda}$ (where $\lambda$ denotes the bare 't Hooft coupling),
 which is parametrically  smaller than $M_Q$ since 
$\lambda\gg1$ in the supergravity regime. The decay into a pair of mesons of this kind can happen when the energy of the static configuration, $2M_Q + V(L)$, equals their total mass, $2M_Q/\sqrt{\lambda}$. Hence
\be
V(L_s) = -2M_Q (1-{1\over \sqrt{\lambda}})\, .
\ee
Due to the presence of the Landau pole, $M_Q$ cannot be as large as we want and it can be interesting to study how $L_s$
varies with the flavor parameters. Relevant plots 
(with $\sqrt{\lambda}=50$) can be found in figure \ref{Ls}. As expected, the screening length decreases as $N_f$ increases and it is an increasing function of $m_q$.
From the Coulomb-like behavior of $V(L)$ we expect, moreover, that $L_s$ is decreasing with $M_Q$. The fact that the potential also depends on the cutoff $M_Q$ (at fixed $L$, $|V(L)|$ is an increasing function of $M_Q$ as we have seen) does not alter this behavior, as it is evident in the figure. 

A recent proposal for improving the quenched approximation in QCD so to take into account the screening effects of dynamical flavors can be found in \cite{armoni}.

\subsection{The string breaking length}  
It is important to notice that due to the smearing, the decay of the static configuration into a given pair of heavy-light bound states is suppressed by $1/N_c$ and not by $N_f/N_c$ as it would happen in a setup with localized parallel flavor branes. 
\begin{figure}[t]
\begin{center}
\includegraphics[width=.3\textwidth]{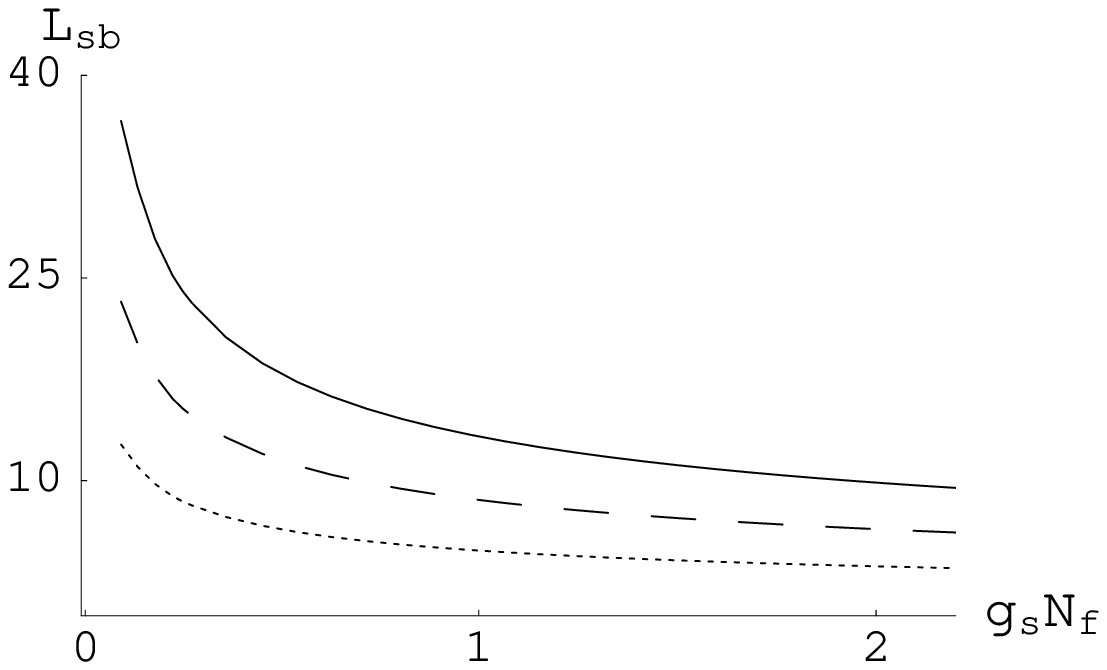}
\includegraphics[width=.3\textwidth]{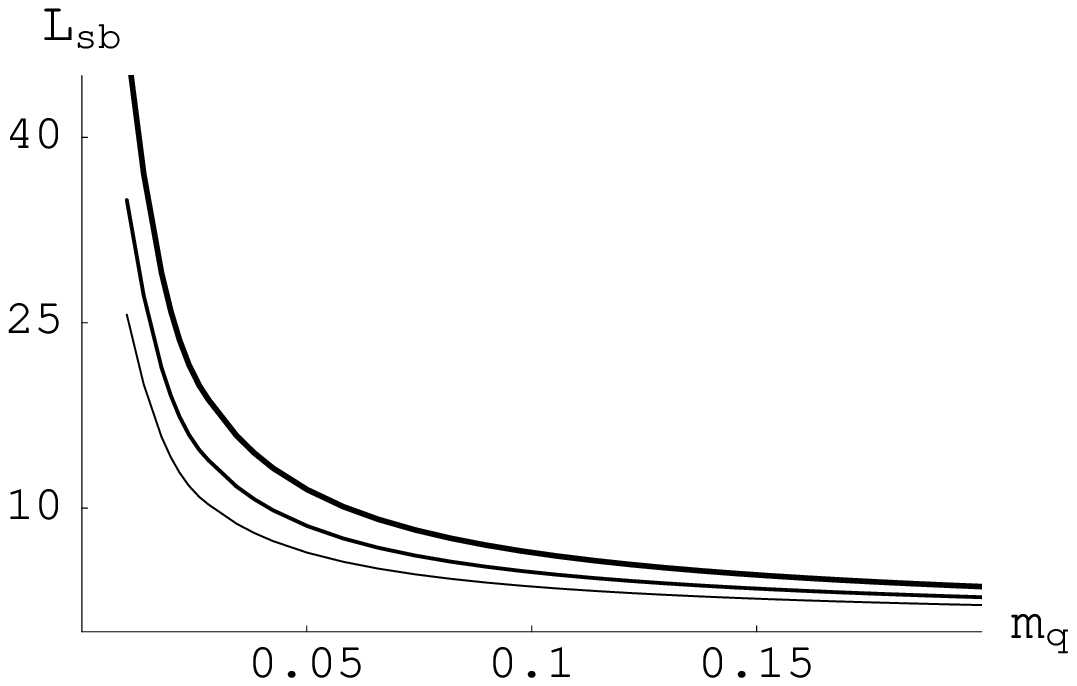}
\includegraphics[width=.3\textwidth]{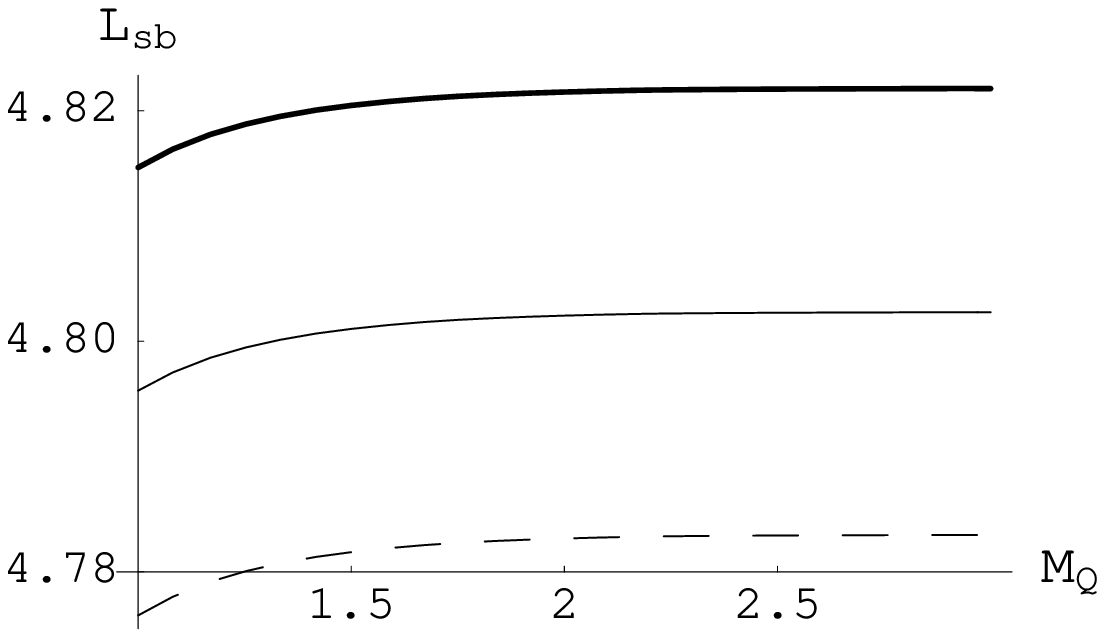}
\end{center}
 \caption{From left to right, the string breaking length $L_{sb}$ as a function of: $g_s N_f$, for $m_q=0.03, 0.05, 0.1$, top to bottom ($M_Q=2$); $m_q$, for $g_s N_f=0.5, 1, 2$, top to bottom ($M_Q=2$); $M_Q$ for $g_sN_f=1, m_q=0.1$ (thick), $g_sN_f=1.01, m_q=0.1$ (solid), $g_s N_f=1, m_q=0.101$ (dashed).}
\label{Lsb1}
\end{figure}
A decay rate which is only $N_f/N_c$ suppressed can be obtained by considering the possibility of decaying into an arbitrary sizable fraction of the $N_f$ types of heavy-light bound states. We call the minimal $\bar Q Q$ separation at which this kind of decay can happen, the ``string-breaking length'' $L_{sb}$. We choose to define $L_{sb}$ as the length at which a dynamical quark-antiquark pair with the same internal charges as the test quarks can be popped out from the vacuum. The corresponding produced heavy-light bound states have a mass given by $M_Q-m_q$ and thus a large binding energy $E_b=-2m_q$ independent on the 't Hooft coupling. The open string describing these ``mesons'' is just stretched along the radial direction of the geometry (\ref{defkwa}) between the probe brane at $\rho_Q$ and a parallel dynamical flavor brane at $\rho_q$ \cite{kkw}. Since the interactions between the two produced ``mesons'' is negligible, it follows from our definition that $L_{sb}$ is just the solution of
\be
V(L_{sb}) + 2M_Q = 2(M_Q-m_q)\quad \Rightarrow \quad V(L_{sb}) = -2m_q\,.
\ee
Figure \ref{Lsb1} 
shows that $L_{sb}$ is a decreasing function of both $N_f$ and $m_q$. 
The latter behavior is due to the shift in the meson mass as we vary $m_q$.
$L_{sb}$ is an increasing function of $M_Q$ which tends to flatten at large $M_Q$.
\subsection{Comparison with the Heaviside approximation}
\begin{figure}[htbp]
 \centering
 \includegraphics[width=.4\textwidth]{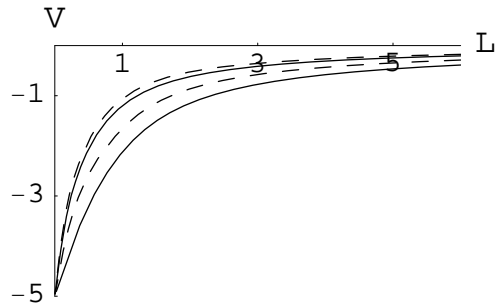}
 \caption{The potential $V(L)$ in the step approximation (dashed lines) and with the correct $N_f(\rho)$ function (solid lines), 
for  $M_Q=2.5$, $g_s N_f=1$ and $m_q=0.3, 1.5$ (top to bottom).}
 \label{compareKW}
 \end{figure}

It is instructive to compare the behavior of the Wilson loop observables as obtained using the correct string dual solution with $g_s N_f(\rho)$ given in (\ref{Nfofr}), with those obtained using the Heaviside step function approximation. In figure \ref{compareKW} we compare the plots for $V(L)$ at different values of the parameters.
The figure confirms that the step approximation, in the flavored KW model, works quite well when $m_q$ is small, or, better said, when $M_Q - m_q$ is large. Moreover, it is possible to show that the approximation is even better if the step is placed where the function $N_f(\rho)$ has reached half of its value $N_f(\rho_{step})=N_f/2$. 

\subsection{Comparison with field theory expectations}
\label{comp}
The behavior of the potential $V(L)$ as well as that of the critical lengths $L_s$ and $L_{sb}$, as a function of the flavor parameters, fits very well with a simple analytical result suggested by field theory and AdS/CFT arguments.
\begin{figure}
\centering
\includegraphics[width=0.4\textwidth]{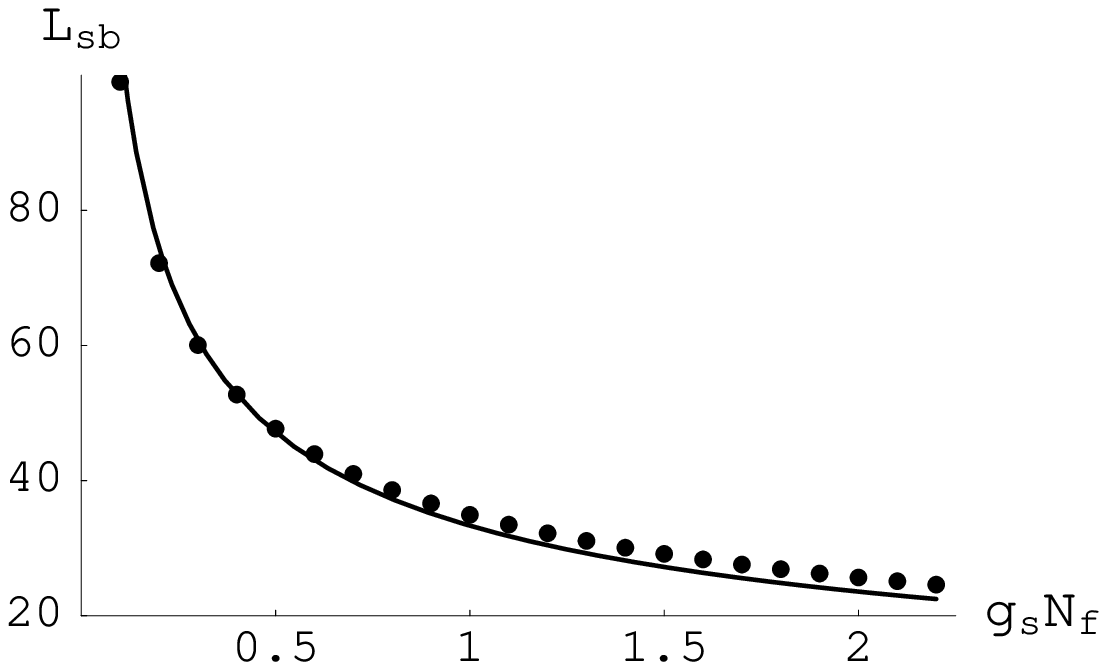} \hfill
\includegraphics[width=0.4\textwidth]{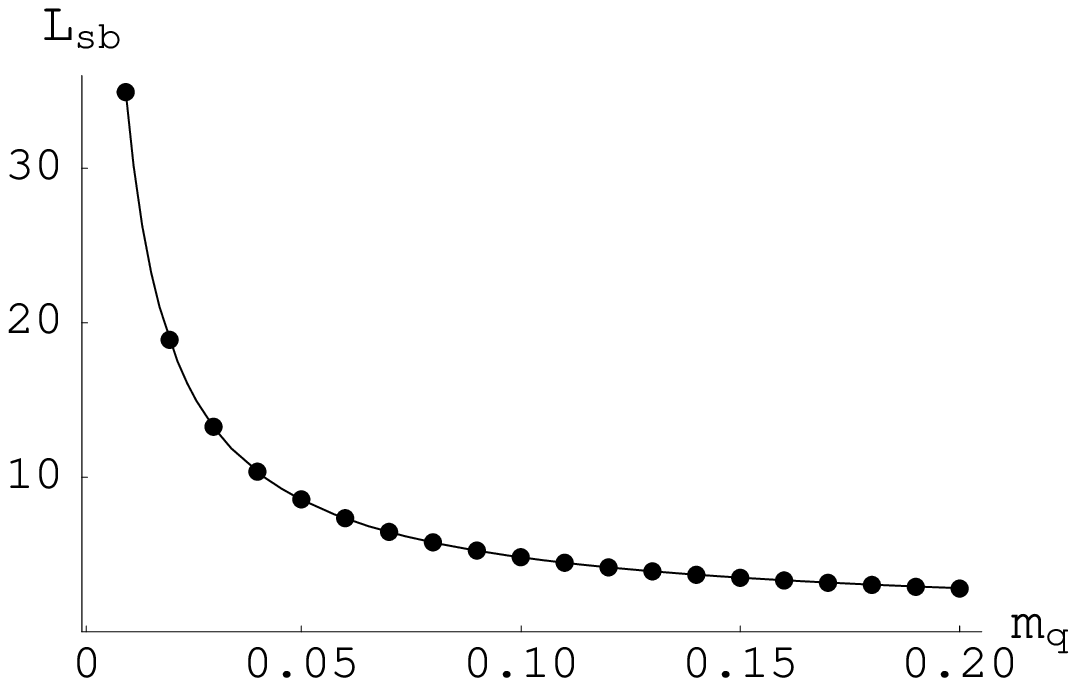}
\caption{Fits of the numerical data (dots) with an analytic formula like (\ref{anL}); on the left, $m_q=0.01, M_Q=2$; on the right, $g_sN_f=1, M_Q=2$.}
\label{fits}
\end{figure}

We know that in the low energy regime, when the massive flavors are integrated out, our model has vanishing beta function. In the string dual description this is reflected into the constancy of the dilaton $\phi=\phi_{IR}$ in the $\rho<\rho_q$ region.  The addition of
flavors gives a negative beta function, whose one loop coefficient behaves as $b_{UV}\approx-N_f$. In the perturbative regime
$m_q\ll\Lambda_{UV}$ (where $\Lambda_{UV}$ indicates the position of the Landau pole), we can match the UV and IR couplings at the scale $m_q$ so to get the IR value of the coupling
\be
{8\pi^2\over g_{IR}^2} \approx N_f \log{\Lambda_{UV}\over m_q}\,\,.
\ee
This expression is nothing but the field theory rewriting of the stringy result we have found for the dilaton in the step function approximation, $e^{\phi_{IR,st}} = - 4\pi/(3g_s N_f \rho_q)$. Indeed, when $\rho_q\rightarrow-\infty$ we can safely identify  $\rho_q\approx\log(m_q/\Lambda_{UV})$. Now, remembering the expression for the $\bar Q Q$ potential of a conformal
theory with an AdS dual \cite{maldawilson}, we expect that for our model, in the strong coupling regime
\be
V_{{\bar Q}Q}(L) \approx -{\sqrt{g_{IR}^2N_c}\over L} \approx - {\sqrt{N_c}\over \sqrt{
  N_f
  \log{\Lambda_{UV}\over m_q}}}\, {1\over L}\,\,,
\label{anV}
\ee
in the large $L$ limit. From this expression we can deduce, in the same limits, the screening length $L_s$ 
\be
L_s \approx {\sqrt{N_c}\over M_Q\sqrt{N_f\log{\Lambda_{UV}\over m_q}}}\,\,,
\label{anLs}
\ee
and the string breaking length $L_{sb}$ 
\be
L_{sb}\approx {\sqrt{N_c}\over m_q\sqrt{N_f\log{\Lambda_{UV}\over m_q}}}\,\,.
\label{anL}
\ee
The above analytical functions share with our numerical results the same qualitative behavior with the flavor parameters. The expression for the screening length is not expected to fit well with our data, since the approximation we used to get eq. (\ref{anLs}) requires $L$ to be large. Instead, in figure \ref{fits} we have fitted our numerical data
with a formula for $L_{sb}$ as in eq. (\ref{anL}) - treating $\Lambda_{UV}$ and an overall constant as parameters - finding a very good agreement.

\vskip 15pt
\centerline{\bf Acknowledgments}
\vskip 10pt
\noindent
We are grateful to Carlos N\'u\~nez for his crucial observations and his participation in the early stages of this project. We would also like to thank D. Are\'an,
R. Argurio, S. D. Avramis, F. Benini, R. Casero, S. Cremonesi, N. Drukker, E. Imeroni, S. Kuperstein, 
A.V. Ramallo, K. Sfetsos, M. Teper, W. Troost. This work has been supported by the European Commission FP6 programme
MRTN-CT-2004 v-005104, ``Constituents, fundamental forces and symmetries in the universe''. F. B. is also supported by the Belgian Fonds de la Recherche
Fondamentale Collective (grant 2.4655.07), by the Belgian Institut
Interuniversitaire des Sciences Nucl\'eaires (grant 4.4505.86) and
the Interuniversity Attraction Poles Programme (Belgian Science
Policy). A. C. is also supported by the FWO -
Vlaanderen, project G.0235.05 and by the Federal Office for
Scientific, Technical and Cultural Affairs through the ‘Interuniversity Attraction
Poles Programme – Belgian Science Policy’ P6/11-P. A. P. is also supported by a NWO VIDI grant 016.069.313 and by
INTAS contract 03-51-6346. F. B. and A. C. thank the Galileo Galilei Institute for Theoretical Physics for hospitality and the INFN for partial support during the completion of this work.

\appendix

\setcounter{equation}{0}
\renewcommand{\theequation}{\Alph{section}.\arabic{equation}}

\section{D7 embeddings from the orbifold}
\label{apporb}
There are two main classes of holomorphic embeddings for D7 branes on the conifold: class I has $z_1=\mu$  \cite{Ouyang:2003df} as representative embedding equation, while class II is represented by $z_1-z_2=\mu$  \cite{kuper}.  The two classes are argued to correspond to two different ways of adding flavors to the KW model. Let us focus here on the massless $\mu=0$ case. Class I embeddings have two branches of D7 branes, each branch adding fundamental matter to the first node of the KW quiver and antifundamental matter to the second. In the class II case a D7-brane provides fundamental and antifundamental matter to one node. Since the KW theory can be obtained starting from the ${\cal N}=2$ conformal theory describing $N_c$ D3-branes on the ${\IC}^2/\IZ_2$ orbifold singularity \cite{kw} it could be useful to classify the possible kinds of D7 embeddings starting from the orbifold picture. This could also help in finding the correct superpotential terms corresponding to the chosen embeddings. A discussion on these issues can be found in  \cite{Ouyang:2003df}.

The orbifold theory is a quiver with gauge group $SU(N_c)\times SU(N_c)$ and, in terms of ${\cal N}=1$ components, the same bifundamental multiplets $A_i, B_i$ as the KW theory, plus two adjoint supermultiplets $\Phi_1$ and $\Phi_2$, one
 for each of the gauge groups. The superpotential of the theory schematically reads
\be
W_{orb}= \Phi_1 (A_1B_1-A_2B_2) - \Phi_2 (B_1A_1-B_2A_2)\,.
\ee
There are mainly two ways of adding massless flavors to both nodes of the quiver: either we add to the above superpotential cubic terms coupling the new flavors to the bifundamentals, i.e. (modulo $SU(2)\times SU(2)$ rotations)
\be
W_{orbI}= W_{orb} + \tilde q_1 (A_1 -A_2) q_2 + \tilde q_2 (B_2-B_1) q_1\,,
\label{wo1}
\ee
or we couple the new flavors with the adjoint supermultiplets
\be
W_{orbII} = W_{orb} + \tilde q_1 \Phi_1 q_1 - \tilde q_2 \Phi_2 q_2\,.
\label{wo2}
\ee
These two possibilities have been considered in the T-dual Type IIA picture \cite{witten97,uranga} to which we refer for the sign conventions. The superpotential in  (\ref{wo1}) reduces supersymmetry by one half, while that in (\ref{wo2}) does not break the ${\cal N}=2$ supersymmetry (the corresponding string model is just an orbifold of the D3D7 system in flat space). In case I each branch of D7 branes extends only along two directions inside the orbifold. In case II, the D7-branes are taken to be extended along the 4 orbifolded directions. It is easy, in this case, to see that the flavored theory has a Landau pole: the one loop coefficient of the beta function (which does not get perturbative corrections beyond one-loop due to the ${\cal N}=2$ supersymmetry) for each group is equal to $b=3N_c-N_c-2N_c-N_f=-N_f$ (where the negative terms $-N_c, -2N_c, -N_f$ are the contributions of the adjoint, bifundamental and fundamental matter superfields, respectively) and on each node the theory is thus IR free.

Let us now go to the ${\cal N}=1$ conifold theory by adding to the above superpotentials the appropriate mass term for the adjoints
\be
W_{m} =  {m_{\Phi}\over 2}(\Phi_1^2-\Phi_2^2)\,.
\ee
Integrating out the adjoint fields in case I, will produce a superpotential of the form
\be
W_I = W_{KW} + h_1\,\tilde q_1 (A_1 - A_2) q_2 + h_2\,\tilde q_2 (B_2-B_1) q_1\,,
\ee
while in case II one will get
\be
W_{II} = W_{KW} +\hat h_1\, \tilde q_1 [A_1B_1-A_2B_2]q_1 + \hat h_2\,\tilde q_2 [B_1A_1-B_2A_2]q_2 + k_1\,(\tilde q_1 q_1)^2 + k_2\,(\tilde q_2 q_2)^2\,.
\ee
Writing the effective mass terms for the flavors in a compact matrix form $\tilde q M q$, the equation $\det M=0$ gives, for the case I, $A_1B_1+A_2B_2-A_1B_2-A_2B_1=0$, i.e. $z_1+z_2-z_3-z_4=0$. This is nothing but one of the possible rotations of the $z_1=0$ embedding: in fact it is our generalized embedding (\ref{genmless}) with $\alpha_3=\alpha_4=-1$. 
In case II, $\det M=0$ gives $A_1B_1-A_2B_2=0$, i.e. the $z_1-z_2=0$ embedding equation.

The massive generalizations of the embeddings above correspond to the addition of standard mass terms $m_1\tilde q_1 q_1 + m_2\tilde q_2 q_2$ to the massless superpotential. In case I the mass terms break the classical flavor symmetry group to $U(N_f)$, while in case II, the $U(N_f)\times U(N_f)$ symmetry is preserved.

It is also tempting to write the class I D7 embeddings using the orbifold coordinates. The $\IC^2/\IZ_2$ orbifold is described by the $xy=z^2$ equation in $\IC^3$. The massless embeddings of class I, should thus be described by an equation like $z=0$, giving rise to the two branches $x=0$ and $y=0$ corresponding to D7 branes extended only over one half of the orbifold directions.

\section {Dependence on the integration constants}
\label{app: integr}
\begin{figure}
 \centering
\includegraphics[width=.4\textwidth]{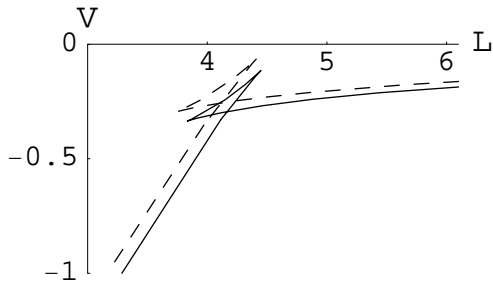}
 \caption{Plots of $V(L)$ with $c_2 = 10$, $m_q=0.2$, $g_sN_f=1$, $M_Q=2.5$. The dashed (resp. solid) line refers to the result obtained using the Heaviside approximation (resp. the correct solution with $N_f(\rho)$).}
\label{Elplots1}
\end{figure}

\setcounter{equation}{0}
In Section \ref{sect: thesol} we have constructed the background solution obtained
with a precise prescription for the integration
constants $\tilde c_1, c_2, c_3$. We would like to briefly discuss here
what would happen with a different choice of constants. As explained in
Section \ref{sec: proposal}, $\tilde c_1 \neq 0$ would produce a singularity in the IR which we
want to avoid. On the other hand, varying $c_3$ just produces a rescaling
of $e^f, e^g, h$ and, for the static quark-antiquark system, a rescaling of $L$ and $V$ which does
not change the qualitative form of the $V(L)$ curves. Finally, $c_2 < 0$
would make the geometry singular at some point below the Landau pole, a
behavior that we also want to avoid. Let us thus focus on modifying our original prescription $c_2=0$ to $c_2>0$. The behavior of the $V(L)$ static potential is crucially affected by the value of $c_2$, as shown in figure \ref{Elplots1}.
For every $c_2>0$ the large $L$ ($\rho_0\to -\infty$) regime of $V(L)$ is unchanged: the potential approaches
zero from below since the theory is IR conformal. However, above a certain critical value of $c_2$ (which depends on the other physical parameters), 
one observes a double turnaround of the $V(L)$ graph at shorter distances: the potential has a discontinuity in the first derivative, and there is a first order
phase transition. 

 This kind of behavior was first found in
\cite{Brandhuber} and afterwards discussed in several different
contexts \cite{doubleturn,screening}. 
In the region of $L$ where $V(L)$ is triple-valued, the point
with lowest $V$ is stable, the intermediate one is metastable and the
one with larger $V(L)$ is perturbatively unstable (always referring to longitudinal perturbations of the string). Stability issues of
this kind of strings were discussed in detail in \cite{Avramis:2006nv}.

The double turnaround region is visible also in the massless $m_q=0$ case.
As it is evident from figure \ref{Elplots1}, the Heaviside approximation does not modify qualitatively the double turnaround behavior of the potential.


\end{document}